\documentclass[showpacs,preprintnumbers,amsmath,amssymb,aps, prd]{revtex4}
\usepackage{graphicx}
\usepackage{epstopdf}
\usepackage{grffile}
\usepackage{caption}
\usepackage{subcaption}
\usepackage{float}
\usepackage{dcolumn}
\usepackage{bm}
\begin{document}
\title{Study of Einstein-bumblebee gravity with Kerr-Sen-like solution in
the presence of a dispersive medium}
\author{Sohan Kumar Jha}
\affiliation{Chandernagore College, Chandernagore, Hooghly-712136
West Bengal, India}
\author{Sahazada Aziz}
\affiliation{Ramananda Centenary
College, Laulara-723151, Purulia, West Bengal, India}
\author{Anisur Rahaman}
\email{anisur.associates@iucaa.ac.in,
manisurn@gmail.com(Corresponding Author)} \affiliation{Durgapur
Govt. College, Durgapur, Burdwan-713214, West Bengal, India}

\date{\today}
\begin{abstract}
\begin{center}
Abstract
\end{center}
A Kerr-Sen-like black hole solution appears in the
Einstein-bumblebee theory of gravity. The solution contains
contains a Lorentz violating parameter in an explicit manner. We
study the null geodesics in the background of this Kerr-Sen-like
black hole surrounded by a dispersive medium like plasma. We
investigate the effect of the charge of the black hole, the
Lorentz violation parameter, and the plasma parameter on the
photon orbits with the evaluation of the effective potential in
the presence of both the Lorentz violation parameter and the
plasma parameter. We also study the influence of the Lorentz
violation parameter and plasma parameter on the emission of energy
from the black hole due to thermal radiation. Besides, we compute
the angle of deflection of massless particles with weak-field
approximation in this generalized situation and examine how it
varies with the Lorentz violation parameter in presence of plasma.
Constraining the parameters of this Lorentz violating
Kerr-Sen-like black hole is also attempted here with the result
obtained from the observations of the Event Horizon Telescope
(EHT) collaboration.

\end{abstract}
\maketitle
\section{Introduction}
The physics of black holes has got renewed interest after the
announcement of the detection of gravitational waves
(GWs)\cite{GWAVE} by the LIGO and VIRGO observatories and the
captured image of the black hole shadow of a super-massive
$\mathrm{M}87^*$ black hole by the EHT based on the very long
baseline interferometry (VIBI) \cite{EHT1, EHT2, EHT3, EHT4, EHT5,
EHT6}. The shadow of black-hole is well studied  from the theory
of General relativity. A black hole shadow is a two-dimensional
dark zone in the celestial sphere caused by the strong gravity of
the black hole. In the article \cite{SYNGE0}, Synge studied the
shadow of the Schwarzschild black hole, which was then termed as
the escape cone of light. The radius of the shadow of the black
hole was calculated in terms of the mass of the black hole and the
radial coordinate where the observer is located. In general, the
shadow of a non-rotating black hole is a standard circle, whereas
it is known that the shadow of a rotating black hole is not a
circular disk. The different features of the black hole have been
widely investigated for various gravity backgrounds, adopting an
almost similar approach based on classical dynamics \cite{KHS,
AMARS, ATAMS, WEIS, ABDUS, FATS, UPAPS, GRENS, ZLIS, STUC, MGAS,
GHOSS, SGGS, GSBS, KUMARS, MANNS, MKHOD1, MKHOD2}.

In General Relativity, light is attributed to the null geodesic of
the spacetime metric and the in-medium effect, which is negligible
for most of the frequency ranges, is significant for the radio
frequency range. In this context, it is worth mentioning the
impact of Solar corona on the time of travel and on the deflection
angle of radio signals that come close to the Sun. Since the
1960s, this influence has been routinely observed. In this case,
one may assume that the medium is a non-magnetized pressureless
plasma and for the gravitational field, the linearized theory is
sufficient. The relevant theoretical development has been made by
Muhleman et al. in the articles \cite{DOM1, DOM2}. With the
available information, the shadow of black holes in presence of
plasma has become an interesting field of research for physicists
since there is a good reason for considering that black holes and
other compact objects are surrounded by plasma. So, naturally,
interest grows to investigate whether the presence of plasma leads
to any observational effect on the radio signals and several
investigations have been carried out in that direction \cite{PS1,
PS2, PS3, PS4, PS5, PS6, PS7, PS8, PS9, FA, LPRO, BABAR}.  The
physics of black holes is crucially linked with the Planck energy
scale where gravity is sufficiently strong and a classical field
theory of curved spacetime is not adequate to capture the finer
details of the Planck scale. It needs quantization of gravity,
which is not well-developed. However, it is reasonable to accept
that the nature of the spacetime is discrete  in that scale.
Lorentz symmetry being a continuous symmetry, would not be tenable
in discrete spacetime. Moreover, according to string theory,
Lorentz invariance should not be an exact principle at all energy
scales \cite{STRING1, STRING2, STRING3, STRING4}. If Lorentz
violation (LV) is considered to be a probe for the foundation of
theoretical physics, then a suitable model is needed, which
contains LV adopted from the standard physical principle. The
known two well-developed and successful theories that describe our
Universe are the theory of General Relativity and the Standard
Model of particle physics. These theories are built up obeying
Lorentz invariance and these two theories are applicable in two
different energy scales. The standard model describes the three
fundamental interactions at the quantum level, whereas general
relativity describes the gravitational interaction at the
classical level. A single complete theory requires the unification
of these two. However, there is a strong obstacle in the way of
unification since gravity can not be quantized in a
straightforward manner. Therefore, a successful unified theory
 has not yet been developed even with huge efforts from different
fronts. To unify them at very high energy scales, one builds an
effective field theory termed as Standard Model Extension (SME),
that couples the Standard Model to the theory of General
Relativity, which involves extra items containing information
about the LV happening at the Plank scale \cite{ESM1, ESM2, ESM3}.
The simplest case is described by a single vector bumblebee field
with a nonzero vacuum expectation value and the spontaneous LV
triggered by a smooth quadratic potential. Bumblebee gravitational
model was first studied by Kostelecky and Samuel in \cite{STRING1,
STRING2, STRING3} as a specific pattern for unprompted Lorentz
violation. Therefore, the Lorentz violation aspect may be
considered as a good probe on the foundation of physics.

In this context, the study of the LV effect on the shadow of a
black hole in the presence of dispersive media like plasma would
be of interest. There are several recent investigations to study
the deformation of black hole shadow due to the LV effect
\cite{CASANA0, CASANA, OUR}. In these studies, the in-medium
(plasma) effect has not been considered so far. Therefore, the
study of the deformation of the shadow of the black hole connected
with Kerr-Sen-like spacetime, the emission of energy from this
black hole due to radiation, and the weak lensing in the presence
of plasma would be interesting extensions. The black hole
associated with Kerr-Sen-like spacetime \cite{OUR} contains an LV
parameter as both the Swarczschild-like and Kerr-like spacetime
contain \cite{CASANA0, CASANA}. This Kerr-Sen-like spacetime will
enable us to study the LV effect in presence of plasma.

It is known that observation of the EHT collaboration has unveiled
the first shadow image of a supermassive black hole
$\mathrm{M}87^*$ with an asymmetric bright emission ring with a
diameter of $42 \pm 3 \mu$$as$ exhibiting a deviation from
circularity $\Delta C\leq 0.10$. It is consistent with the shadow
of the Kerr black hole, but the quantitative features are not
always sufficient to distinguish between black holes predicted
from different theories of gravity. Thus, there is enough room for
investigation concerning the viability of black holes predicted
from modified theories of gravity with the help of the observed
data of $\mathrm{M}87^*$ black hole shadow by the EHT
collaboration placing constraints on the black hole parameters,
and consequently, it becomes an important tool to test the
phenomena of strong gravitational fields. Extensive development in
this direction is going on \cite{PVP, EHT, MOHAN, VOL, PAP, MISBA,
SAFIQ}.

Therefore, in this paper, we will consider Einstein's gravity,
which is coupled to the bumblebee field to get some suppressed
effects emerging from the underlying unified quantum gravity
theory on our low energy scale. Recently, Casana et al.
\cite{CASANA0} gave an exact Schwarzschild-like solution of this
bumblebee gravity model, and considered some classical tests. The
rotating black hole solutions are the most relational subsets of
astrophysics. In \cite{CASANA}, Ding et al. found an exact
Kerr-like solution by solving Einstein-bumblebee gravitational
field equations and studied its black hole shadow. Later on, in
the article \cite{LIALI}, the authors have studied the weak
gravitational deflection angle of relativistic massive particles
by this Kerr-like black hole. In \cite{OUR}, we have presented a
Kerr-Sen-like modified solution. It has an LV parameter and the
presence of plasma supplies an additional parameter. In this
article along with the study of theoretical aspects concerning the
Kerr-Sen-like black hole in the presence of plasma, we have made
an attempt to constrain the parameters of Kerr-Sen-like modified
black holes considering $\mathrm{M}87^*$ as Kerr-Sen-black holes
using the recently obtained result of the EHT collaboration.

The article is organized as follows. In Sec. II, we study the null
geodesics around the charged rotating Kerr-Sen-like black hole in
the presence of plasma. In Sec. III, we study the shadow
corresponding to this charged rotating black hole and how the
shadow gets deformed with the variation of charge, LV parameter,
and the plasma parameter. Sec. IV is devoted to the study of the
emission of energy due to thermal radiation for this black hole,
and the study of the effect of the LV parameter and the plasma
parameter on it. In Sec. V, we study the deflection angle of light
with weak-field approximation in this LV spacetime background in
the presence of plasma. Sec. VI is devoted to constraining the
parameters of the Kerr-Sen-like black hole with the experimentally
observed quantities for the $\mathrm{M}87^{*}$ black hole. The
final Sec. VII contains a summary and discussion.

\section{Null geodesics around the charged rotating Kerr-Sen-like
black hole in the presence of plasma} It is known that
Einstein-bumblebee gravity model is a modification over the
Einstein's gravity where a pseudo vector field namely bumblebee
field was introduced that broke Lorentz symmetry, acquiring a
vacuum expectation value. In the article \cite{CASANA0}, the
authors showed that this modified theory provided for a
Schwarzschild-like solution. A Kerr-like solution was developed in
\cite{CASANA}. We have shown that Einstein-bumblebee gravity model
also renders a Kerr-Sen-like black hole metric. It is the outcome
of the introduction of the Bumblebee field in the Kerr-Sen
background \cite{ASEN} In Boyer-Lindquist coordinates, the
Kerr-Sen-like charged rotating metric that results out from the
Einstein-bumblebee gravity reads \cite{OUR}
\begin{eqnarray}
d s^{2}=-\left(1-\frac{2 M r}{\rho^{2}}\right) d t^{2}-\frac{4 M r
a \sqrt{1+l} \sin ^{2} \theta}{\rho^{2}} d t d
\varphi+\frac{\rho^{2}}{\Delta} d r^{2}+\rho^{2} d
\theta^{2}+\frac{A \sin ^{2} \theta}{\rho^{2}} d \varphi^{2},
\label{METRIC}
\end{eqnarray}
where
\begin{eqnarray}
\rho^{2}=r\left(r+b\right)+a^{2}\left(1+l\right),
\Delta=\frac{r(r+b)-2 M r}{1+l}+a^{2}, A=\left[r(r+b)+(1+l)
a^{2}\right]^{2}-\Delta(1+l)^{2} a^{2} \sin ^{2} \theta.
\end{eqnarray}
The parameters $a$ and $b$ are related with angular momentum and
the charge of the black hole by the relations $a=\frac{J}{M}$ the
$Q=\sqrt{bM}$. Here $M$, $Q$ and $J$ respectively refer to the
mass, charge and angular momentum. The expression of $\rho$ and
$\Delta$ contain the parameter $l$, that represents the Lorentz
violation parameter, which is associated with spontaneous
violation of symmetry of the vacuum of the bumblebee field. If $l
\rightarrow 0 $ it recovers the usual Kerr-Sen metric and for $a
\rightarrow 0$ and $b \rightarrow 0$ the metric turns into
\begin{eqnarray}
d s^{2}=-\left(1-\frac{2 M}{r}\right) d t^{2}+\frac{1+l}{1-2 M /
r} d r^{2}+r^{2} d \theta^{2}+r^{2} \sin ^{2} \theta d
\varphi^{2}, \end{eqnarray}
which is the Schwarzschild-like
solution for the Einstein-bumblebee gravity \cite{CASANA0,
CASANA}.

We consider a static inhomogeneous plasma in the gravitational
field with a refractive index $n$. The expression of refractive
index $n$, as formulated by Synge \cite{SYNGE}, in terms of
dynamical variables reads
\begin{equation}
n^{2}=1+\frac{p_{\mu} p^{\mu}}{\left(p_{\nu}
u^{\nu}\right)^{2}},\label{R.I}
\end{equation}
where $p_{\mu}$ and $u^{\nu}$ are four-momentum and four-velocity
of the massless particle respectively. Let us now start with the Hamiltonian
for massless particles like photon:
\begin{equation}
H\left(x^{\mu}, p_{\mu}\right)=\frac{1}{2}\left[g^{\mu \nu}
p_{\mu} p_{\nu} -\left(n^{2}-1\right)\left(p_{0}
\sqrt{-g^{00}}\right)^{2}\right].
\end{equation}
The standard definitions $x^{\mu}=\partial H / \partial p_{\mu}$,
and $\dot{p_{\mu}}=\partial H / \partial x^{\mu}$ lead us to write
down the equations of motion for photon in the plasma medium as
follows
\begin{eqnarray}
\rho^{2} \frac{d r}{d \lambda}&=&\pm \sqrt{R},\quad \rho^{2} \frac{d
\theta}{d \lambda}=\pm \sqrt{\Theta},\\
(1+l) \Delta \rho^{2} \frac{d t}{d \lambda}&=&An^{2}-2
\sqrt{1+l} Mra\xi,\\
(1+l) \Delta \rho^{2} \frac{d \phi}{d \lambda}& =&2M r a
\sqrt{1+l}+\frac{\xi}{\sin ^{2} \theta}\left(\rho^{2}-2 M
r\right),
\end{eqnarray}
where $\lambda$ is the affine parameter and
\begin{eqnarray}\nonumber
R(r)&=&\left[\frac{r(r+b)+(1+l) a^{2}}{\sqrt{1+l}}-a\xi\right]^{2}
-\Delta\left[\eta+(\xi-\sqrt{1+l}a)^{2}\right]+\left(n^{2}-1 \right)
\frac{\left[r(r+b)+a^{2}(1+l)\right]^{2}}{1+l},\\
\Theta(\theta)&=&\eta+(1+l) a^{2} \cos ^{2} \theta-\xi^{2} \cot
^{2} \theta -\left(n^{2}-1 \right)a^{2}\left(1+l
\right)sin^{2}\theta. \label{CONP}
\end{eqnarray}
In equation (\ref{CONP}), we introduce two conserved
parameters $\xi$ and $\eta$ as usual, which are defined by
\begin{eqnarray}
\xi=\frac{L_{z}}{E} \quad \textrm{and} \quad \eta=\frac{\mathcal{Q}}{E^{2}},
\end{eqnarray}
where $E, L_{z}$ and $\mathcal{Q}$ are the energy, the axial
component of the angular momentum, and the $Carter~constant$
respectively.

We introduce a static inhomogeneous plasma with a refractive index
$n$, which depends on the space location $x$, and the photon
frequency $\omega$ \cite{FA}. In terms of $\omega$, the square of
the refractive index $n$ is defined by
\begin{equation}
n^{2}=1-\frac{\omega_{e}^{2}}{\omega^{2}},
\end{equation}
where $\omega$ and $\omega_{e}$ are, respectively, the frequency of
photon and the electron-plasma frequency. In the general theory of
relativity, the redshift scenario entails that frequency of
photons depends on the spatial coordinates due to the presence of
the gravitational field. Besides, the electron-plasma frequency
has the expression \cite{AA}
\begin{equation}
\omega_{e}^{2}=\frac{4 \pi e^{2} N(r)}{m},
\end{equation}
where $N(r)$ is the concentration of electron in the inhomogeneous
plasma. The mass and the charge of the electron are respectively
denoted by $m$ and $e$. We now consider a radial power-law
density as
\begin{equation}
N(r)=\frac{N_{0}r_{0}}{r^{h}},
\end{equation}
where $N_{0}$ is the density number at the radial position of the
inner edge of plasma environment $r_{0}$.
Therefore, we have
\begin{equation}
\omega_{e}^{2}=\frac{4 \pi e^{2} N_{0}r_{0}}{m r^{h}},
\end{equation}
where $h \geq 0$. Here we consider $h=1$, as it has been
considered in the article \cite{AR}, so that we have
\begin{equation}
n=\sqrt{1-\frac{k}{r}}.
\end{equation}
With this specific form of refractive index, we study various
aspects of photon geodesics with the spacetime metric given in
Eqn. (\ref{METRIC}). The radial equation of motion has the known
form
\begin{eqnarray}
\left(\rho^{2} \frac{d r}{d \lambda}\right)^{2}+V_{e f f}=0.
\end{eqnarray}
The effective potential $V_{e f f}$ in this situation reads
\begin{eqnarray}
V_{e f f}=-[\frac{r(r+b)+(1+l) a^{2}}{\sqrt{1+l}}-a
\xi]^{2}+\Delta\left[\eta+(\xi-\sqrt{1+l}a)^{2}\right]
-\left(n^{2}-1 \right)\frac{\left[r(r+b)+a^{2}(1+l)\right]^{2}}{1+l}.
\end{eqnarray}
Note that it contains both factors $l$ and $k$. The shape of
the orbit crucially depends on the nature of the effective
potential. Therefore, it is natural that the orbit will have crucial
dependence on both factors $l$ and $k$. The
unstable spherical orbit on the equatorial plane will be obtained
if the following conditions are met:
\begin{eqnarray}
\theta=\frac{\pi}{2},\quad R(r)=0,\quad \frac{d R}{d r}=0,\quad \frac{d^{2} R}{d
r^{2}}>0,\quad \eta=0. \label{CONDITION1}
\end{eqnarray}
We plot $V_{e f f}$ versus $r/M$ with $\xi=\xi_{c}+0.2$, where
$\xi_{c}$ is the value of $\xi$ for equatorial spherical unstable
direct orbit. In Fig. 1, we find that the turning point moves
to the left with the increase of the value of $b$ for fixed values
of $k$ and $l$. It also shifts to the left with the increasing
value of $k$ when $b$ and $l$ remain fixed. For the positive value
of the Lorentz violating parameter $l$, it also shifts to the
left and for the negative value of the $l$ it gets shifted
towards the right like the case when plasma was absent \cite{OUR}.
In Fig. 2, if we observe the plot of the critical radius we find
that $r_c$ decreases with the increase of the value of $b$ for
fixed $k$ and $l$; and it increases with the increase of the value
of $k$ when $b$ and $l$ remain fixed. However, for $l
> 0$ the critical radius $r_c$ decreases and for $l< 0$ it
increases.

\begin{figure}[H]
\centering
\begin{subfigure}{.58\textwidth}
\includegraphics[width=.8\linewidth]{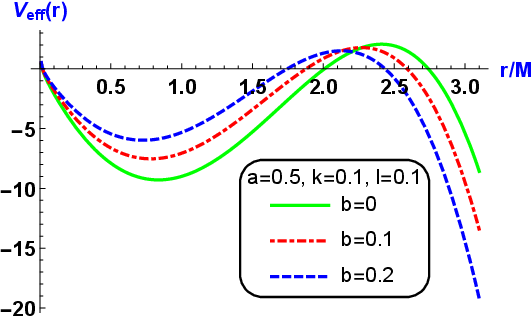}
\caption{Variation of the  effective potential ($V_{eff}$) with
respect to\\ $r/M$ for various values of $b$ with $a=0.5, l=0.1,
k=0.1 $, and $\theta=\pi/2$.}
\end{subfigure}%
\begin{subfigure}{.58\textwidth}
\includegraphics[width=.8\linewidth]{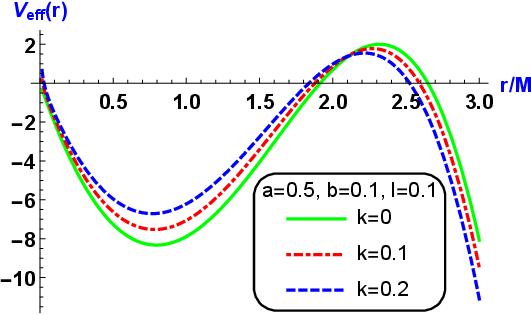}
\caption{Variation of the effective potential ($V_{eff}$) with
respect to\\ $r/M$ for various values of $k$ with $a/M=0.5, l=0.1,
b=0.1$, and $\theta=\pi/2$.}
\end{subfigure}
\par\bigskip
\begin{subfigure}{.58\textwidth}
\centering
\includegraphics[width=.8\linewidth]{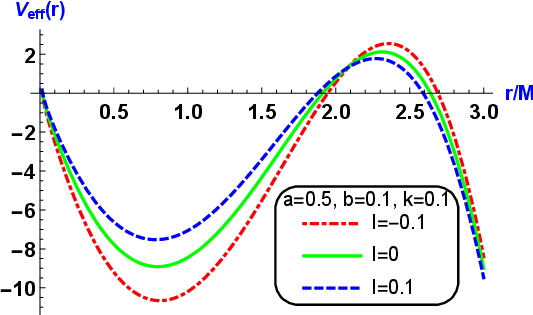}
\caption{Variation of effective potential ($V_{eff}$) with respect
to\\ $r/M$ for various values of $l$ with $a/M=0.5, k=0.1, b=0.1$,
and $\theta=\pi/2$.}
\end{subfigure}
\caption{Plots of the effective potential for various scenarios.}
\end{figure}
We also plot critical radius versus $a$ keeping
$\theta=\frac{\pi}{2}$. To plot it, we consider variation of the
parameters $b$, $k$ and $l$ involved in our proposed model.
\begin{figure}[H]
\centering
\begin{subfigure}{.58\textwidth}
\includegraphics[width=.8\linewidth]{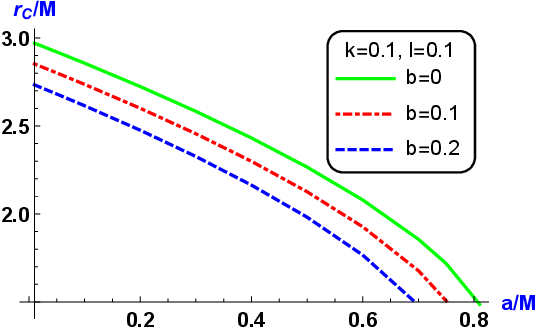}
\caption{Variation of the critical radii $r_c$ with respect to
$a/M$ for various\\ values of $b$ with $ l=0.1, k=0.1$, and
$\theta=\pi/2$.}
\end{subfigure}
\begin{subfigure}{.58\textwidth}
\includegraphics[width=.8\linewidth]{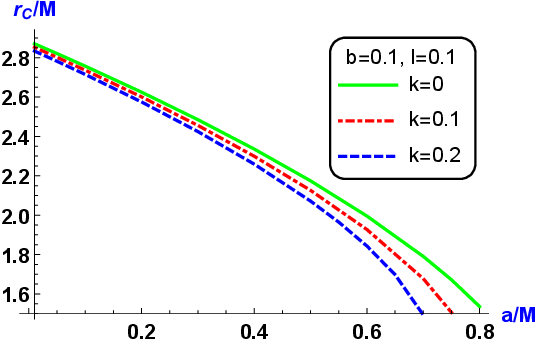}
\caption{Variation of the critical radii  $r_c$ with respect to
$a/M$ for various\\ values of $k$ with $l=0.1, b=0.1$, and
$\theta=\pi/2$.}
\end{subfigure}
\par\bigskip
\begin{subfigure}{.58\textwidth}
\centering
\includegraphics[width=.8\linewidth]{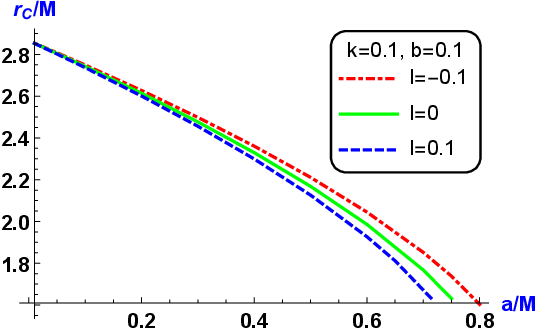}
\caption{Variation of the critical radii  $r_c$ with respect to
$a/M$ for various\\ values of $l$ with $k=0.1, b=0.1$, and
$\theta=\pi/2$.}
\end{subfigure}
\caption{Plots of Critical radius for various scenarios.}
\end{figure}

For more generic spherical orbits where $\theta \neq \pi / 2$ and
$\eta \neq 0$, the conserved quantities $\xi_{s}$ and $\eta_{s}$
for $ r=r_{s}$, are given by the simultaneous solutions of the
equations
\begin{eqnarray}
R=0 \quad \textrm{and} \quad \frac{dR}{dr}=0.
\end{eqnarray}
where
\begin{eqnarray}
R(r)&=&\left[\frac{r(r+b)+(1+l) a^{2}}{\sqrt{1+l}}-a\xi\right]^{2}
-\Delta\left[\eta+(\xi-\sqrt{1+l}a)^{2}\right]+\left(n^{2}-1
\right) \frac{\left[r(r+b)+a^{2}(1+l)\right]^{2}}{1+l}
 \label{RR}
\end{eqnarray}
and
\begin{eqnarray}\nonumber
\frac{dR}{dr}&=&2\frac{2r+b}{\sqrt{1+l}}\left[\frac{r(r+b)+(1+l) a^{2}}{\sqrt{1+l}}-a\xi\right]
-\frac{2r+b-2M}{1+l}\left[\eta+(\xi-\sqrt{1+l}a)^{2}\right]+\frac{k}{r^{2}}
\frac{\left[r(r+b)+a^{2}(1+l)\right]^{2}}{1+l}\\&&-2\frac{k}{r\left(1+l\right)}\left(2r+b\right)\left[r(r+b)+a^{2}(1+l)\right]
\end{eqnarray}
The  conserved quantities $\xi_{s}$ and $\eta_{s}$ are obtained by
solving the equations
\begin{eqnarray}
R\mid_{r=r_s}=0 \quad \textrm{and} \quad
\frac{dR}{dr}\mid_{r=r_s}=0,
\end{eqnarray}
which give
\begin{equation}
\begin{aligned}
\xi_{s}=&-\frac{1}{a^{2} \sqrt{1+l}(b-2 M+2 r)} \left(2 a^{3} M+2 a^{3} l M-2 a M r^{2}+\sqrt{1+l}\right(\frac { a ^ { 2 }} { ( 1 + l ) r ^ { 2 } } \left(4 M^{2} r^{2}\left(-a^{2}(1+l)+r^{2}\right)^{2}+\right. \\
&(b-2 M+2 r)\left(a^{2}(1+l)+r(b+r)\right)\left(a^{4} k(1+l)^{2}+r^{3}\left(b^{2}+4 k M-b(k+2 M-3 r)-\right.\right. \\
&\left.\left.\left.(k+6 M) r+2 r^{2}\right)+a^{2}(1+l) r(-4 k
M+b(k+r)+2 r(M+r))\right)\right)\Big)^{1/2}\bigg)
\end{aligned}
\end{equation}
\begin{equation}
\begin{aligned}
\eta_{s}=&\frac{1}{a^{3}(1+l)^{3 / 2} r(b-2 M+2 r)^{2}}\left(-a^{5} k(1+l)^{5 / 2}(b-2 M+2 r)(2 b-4 M+3 r)-\right.\\
&2 a^{3}(1+l)^{3 / 2} r\left(b k(b-2 M)^{2}+4 b k(b-2 M) r+(5 b k-2(b+k) M) r^{2}+2(k-2 M) r^{3}\right)\\
&-a r^{3}\sqrt{1+l}\left(b(b-2 M)\left(b^{2}+4 k M-b(k+2 M)\right)+2\left(3 b^{3}-4 k M^{2}+8 b M(k+M)-2 b^{2}(k+5 M)\right) r+\right.\\
&\left.\left(13 b^{2}+10 M(k+2 M)-b(5 k+32 M)\right) r^{2}+2(6 b-k-8 M) r^{3}+4 r^{4}\right)+\\
&4(1+l) M r^{3} \left(\frac{a^{8} k(1+l)^{2}(b-2 M+2 r)}{r^{2}}+\right.\\
&\frac{a^{6}(1+l)\left(2 k(b-2 M)^{2}+\left(b^{2}+5 b k-10 k M\right) r+2(2 b+k) r^{2}+4 r^{3}\right)}{r}\\
&+\frac{a^{2} r^{2}}{1+l}\left(b(b-2 M)\left(b^{2}+4 k M-b(k+2 M)\right)+2\left(3 b^{3}-4 k M^{2}+8 b M(k+M)-2 b^{2}(k+5 M)\right) r+\right.\\
&\left.\left(13 b^{2}+2 M(5 k+8 M)-b(5 k+32 M)\right) r^{2}+2(6 b-k-8 M) r^{3}+4 r^{4}\right)+\\
&a^{4}(b k(b-4 M)(b-2 M)+2 b(b(b+k)-2(b+2 k) M) r+\\
&\left.\left(10 b^{2}+2 k M-b(k+16 M)\right) r^{2}+2(8 b-k-8 M) r^{3}+8 r^{4})\bigg)^{1/2}\right)
\end{aligned}
\end{equation}

\section{BLACK HOLE SHADOW}
To describe the nature of the shadow, that an observer see in
the sky, the following two celestial coordinates will be helpful
\begin{eqnarray}\nonumber
\alpha(\xi, \eta ; \theta)&=& -\frac{\xi_{s} \csc
\theta}{n},\\\nonumber \beta(\xi, \eta ;
\theta)&=&\frac{\sqrt{\eta+(1+l) a^{2} \cos ^{2} \theta-\xi^{2}
\cot
^{2} \theta -\left(n^{2}-1 \right)a^{2}\left(1+l \right)sin^{2}\theta}}{n}.\\
\end{eqnarray}
Let us now proceed to see the shape and size of the shadows. The
nature of the shadow will depend on various parameters. So, we
sketch the black hole shadow for various possible cases. In Fig. 3 and Fig. 4
we see that with the increase in the value of $b$ the
size of the shadow decreases and the left end of the shadow moves
a little to the right when $k$ and $l$ remain fixed.  Fig. 5 and Fig. 6 show the change
of shape of the shadow with a variation of the plasma parameter
$k$ keeping $b$ and $l$ fixed. The figure shows that the size of
the shadow increases with the increase of the value of $k$. Here
deformation of the shape of the shadow can not be found
pictorially however mathematically very little deviation is
observed, which can be observed from TABLE. I. Fig. 7 and Fig. 8 show that for a negative
value of $l$ the left end of the shadow shifts towards the left
and for a positive value of $l$ it shifts towards the right.
Deformation of the shape of the shadow is prominent in all cases.
\begin{figure}[H]
\centering
\begin{subfigure}{.58\textwidth}
\includegraphics[width=.8\linewidth]{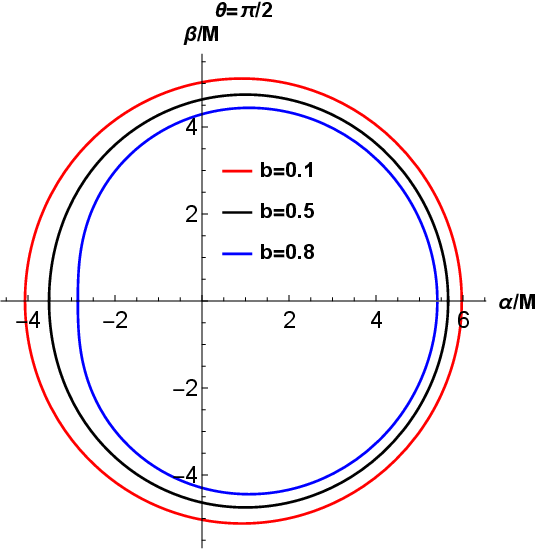}
\end{subfigure}%
\begin{subfigure}{.58\textwidth}
\includegraphics[width=.8\linewidth]{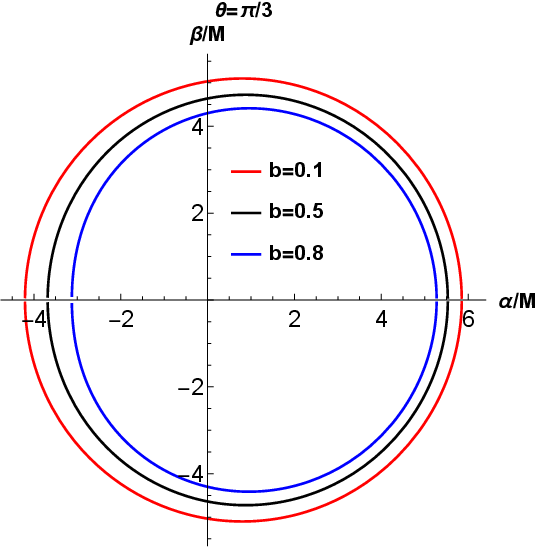}
\end{subfigure}
\caption{The figure of the shadows for different values of $b$
when $\theta=\frac{\pi}{2}$ and $\theta=\frac{\pi}{3}$ with
$a=0.4, k=0.2$, and $l=0.2$.}
\end{figure}

\begin{figure}[H]
\centering
\begin{subfigure}{.58\textwidth}
\includegraphics[width=.8\linewidth]{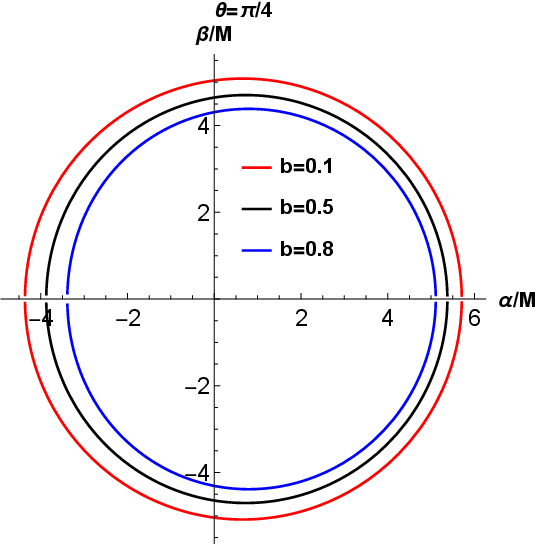}
\end{subfigure}%
\begin{subfigure}{.58\textwidth}
\includegraphics[width=.8\linewidth]{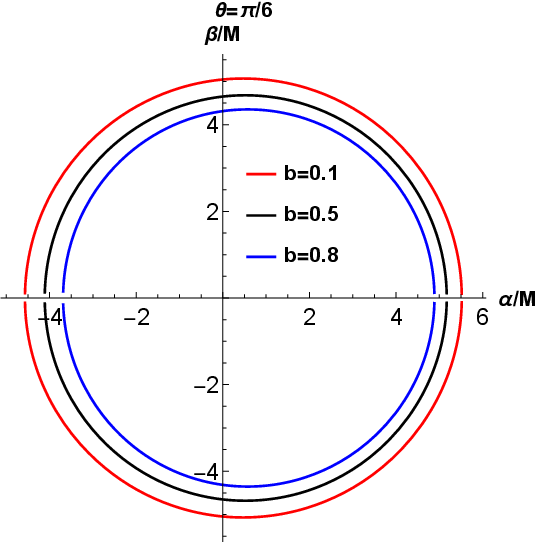}
\end{subfigure}
\caption{The figure of the shadows for different values of $b$
when $\theta=\frac{\pi}{4}$ and $\theta=\frac{\pi}{6}$ with
$a=0.4,k=0.2$, and $l=0.2$.}
\end{figure}

\begin{figure}[H]
\centering
\begin{subfigure}{.58\textwidth}
\includegraphics[width=.8\linewidth]{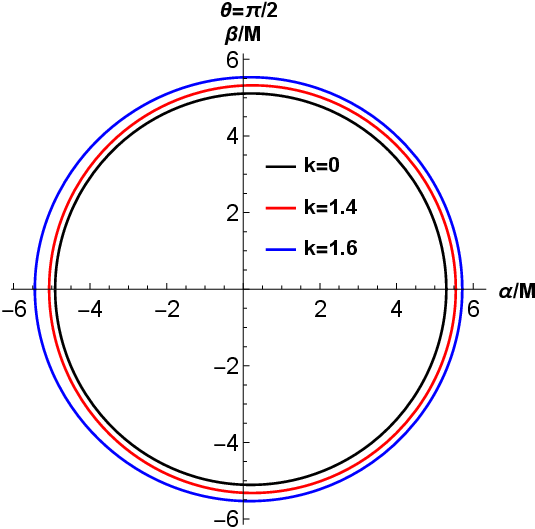}
\end{subfigure}%
\begin{subfigure}{.58\textwidth}
\includegraphics[width=.8\linewidth]{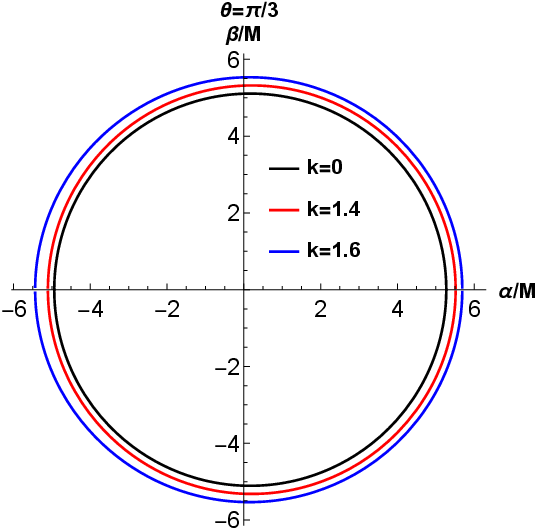}
\end{subfigure}
\caption{The figure of the shadows for different values of $k$
when $\theta=\frac{\pi}{2}$ and $\theta=\frac{\pi}{3}$ with
$a=0.1, b=0.1$, and $l=-0.1$.}
\end{figure}

\begin{figure}[H]
\centering
\begin{subfigure}{.58\textwidth}
\includegraphics[width=.8\linewidth]{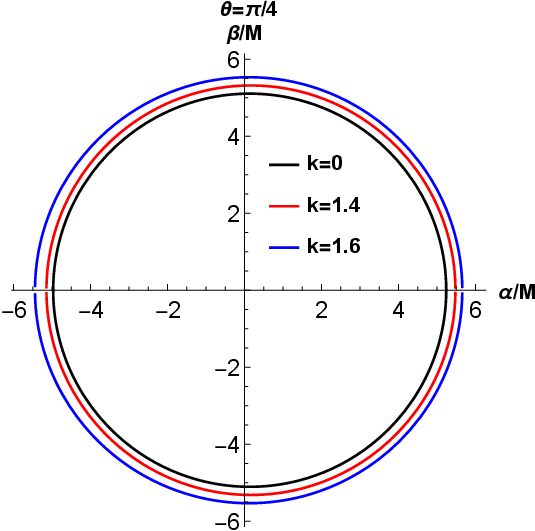}
\end{subfigure}%
\begin{subfigure}{.58\textwidth}
\includegraphics[width=.8\linewidth]{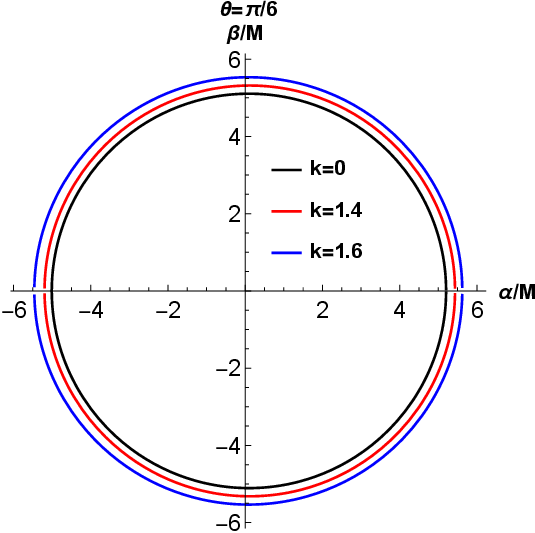}
\end{subfigure}
\caption{The figure of the shadows for different values of $k$
when $\theta=\frac{\pi}{4}$ and $\theta=\frac{\pi}{6}$ with
$a=0.1, b=0.1$, and $l=-0.1$.}
\end{figure}

\begin{figure}[H]
\centering
\begin{subfigure}{.58\textwidth}
\includegraphics[width=.8\linewidth]{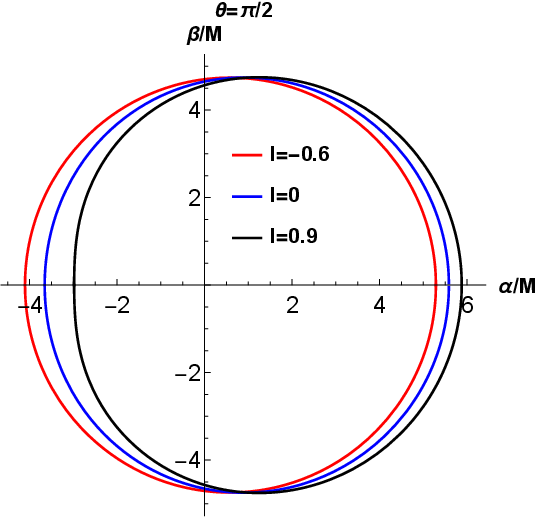}
\end{subfigure}%
\begin{subfigure}{.58\textwidth}
\includegraphics[width=.8\linewidth]{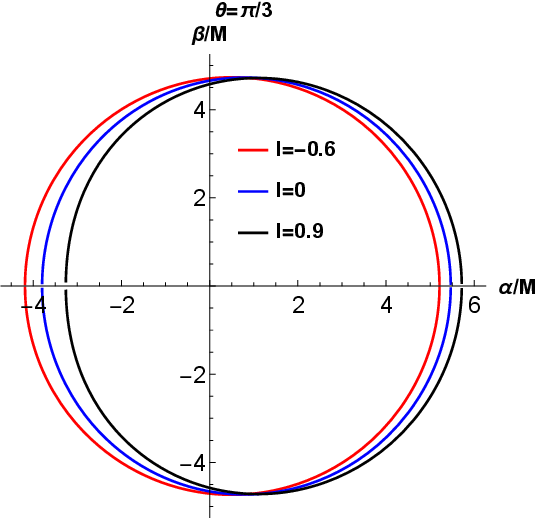}
\end{subfigure}
\caption{The figure of the shadows for different values of $l$
when $\theta=\frac{\pi}{2}$ and $\theta=\frac{\pi}{3}$ with
$a=0.4, b=0.5$, and $k=0.2$.}
\end{figure}

\begin{figure}[H]
\centering
\begin{subfigure}{.58\textwidth}
\includegraphics[width=.8\linewidth]{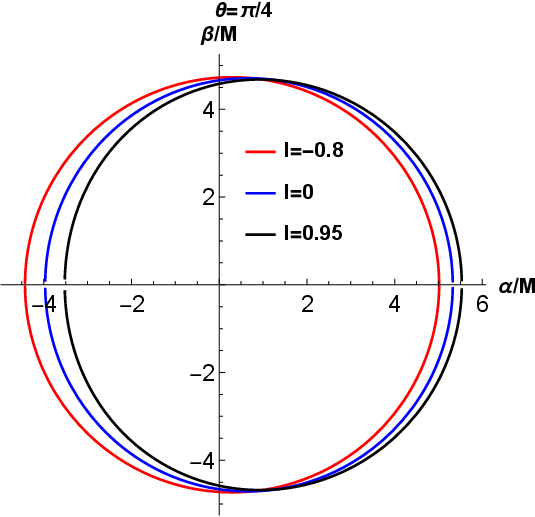}
\end{subfigure}%
\begin{subfigure}{.58\textwidth}
\includegraphics[width=.8\linewidth]{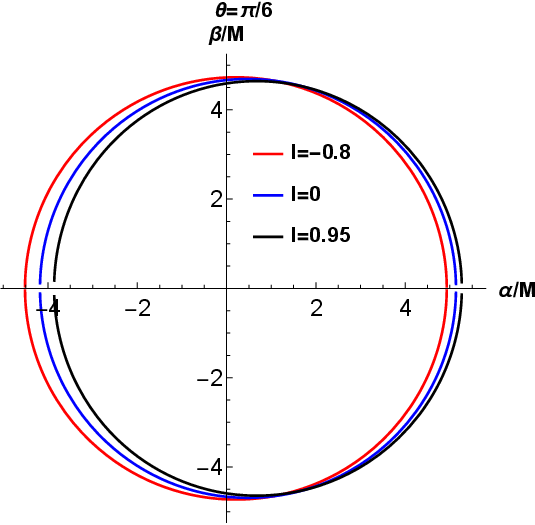}
\end{subfigure}
\caption{The figure of the shadows for different values of $l$
when $\theta=\frac{\pi}{4}$ and $\theta=\frac{\pi}{6}$ with
$a=0.4, b=0.5$, and $k=0.2$.}
\end{figure}
Using the parameters, which are introduced by Hioki and Maeda in
\cite{KHS}, we analyze the deviation from the circular form of the
shadow $\left(\delta_{s}\right)$ and the size
$\left(R_{s}\right)$ of the shadow image of the black hole.
\begin{figure}[H]
\centering
\begin{subfigure}{.58\textwidth}
\includegraphics[width=.8\linewidth]{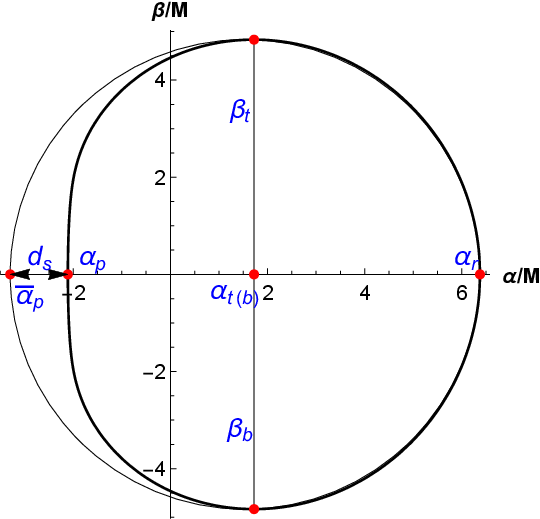}
\end{subfigure}
\caption{A sketch of the black hole shadow and the reference
circle. The distance between the extreme left point of the shadow,
and the reference circle is represented by $d_{s}$.}
\end{figure}
For calculating these parameters, we consider five points. The
four points $\left(\alpha_{t}, \beta_{t}\right),
\left(\alpha_{b}, \beta_{b}\right), \left(\alpha_{r},
0\right)$ and $\left(\alpha_{p}, 0\right)$ are, respectively, the top, bottom, rightmost,
and the leftmost point of the
shadow; and $\left(\bar{\alpha}_{p}, 0\right)$ is the leftmost
point of the reference circle. From simple geometry, we have expressions for the radius $R_{s}$ and the deviation from
circularity $\delta_{s}$ as follows
\begin{equation}
R_{s}=\frac{\left(\alpha_{t}-\alpha_{r}\right)^{2}
+\beta_{t}^{2}}{2\left(-\alpha_{t}+\alpha_{r}\right)},\label{RS}
\end{equation}
and
\begin{equation}
\delta_{s}=\frac{\left(-\bar{\alpha}_{p}+\alpha_{p}\right)}{R_{s}}.\label{DELTAS}
\end{equation}
For all subsequent plots, we have taken $\theta=\pi/2$.

\begin{figure}[H]
\centering
\begin{subfigure}{.58\textwidth}
\includegraphics[width=.8\linewidth]{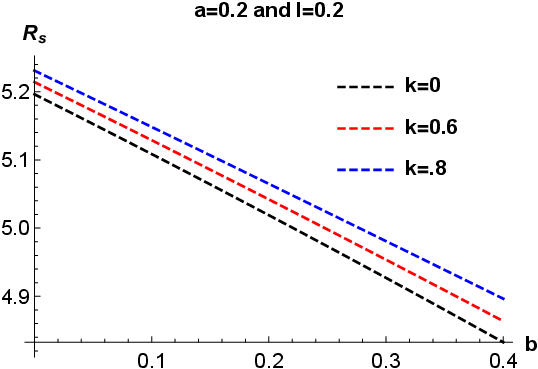}
\end{subfigure}%
\begin{subfigure}{ .58\textwidth}
\includegraphics[width=.8\linewidth]{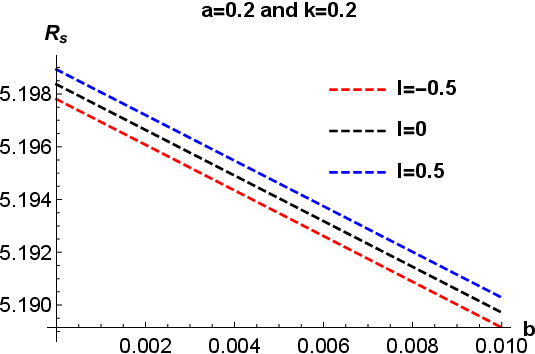}
\end{subfigure}
\caption{The  figure at the left side represents the variation of
shadow radius $R_{s}$ with $b$ for different values of $k$ with
$a=0.2$ and $l=0.2$. The  figure at the right side represents the
variation of $R_{s}$ with $b$ for different values of $l$ with
$a=0.2$ and $k=0.2$.}
\end{figure}

\begin{figure}[H]
\centering
\begin{subfigure}{ .58\textwidth}
\includegraphics[width=.8\linewidth]{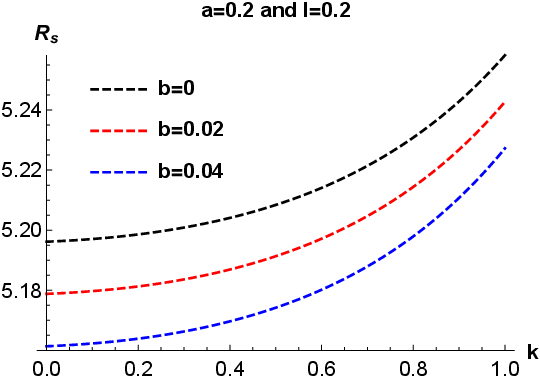}
\end{subfigure}%
\begin{subfigure}{ .58\textwidth}
\includegraphics[width=.8\linewidth]{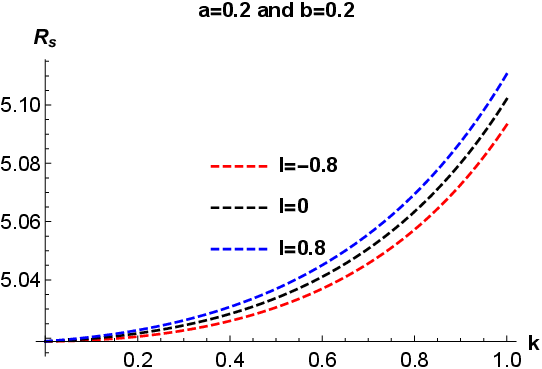}
\end{subfigure}
\caption{The left one represents the variation of shadow radius $R_{s}$ with $k$
for different values of $b$ with $a=0.2$ and $l=0.2$. The
right one represents the variation of $R_{s}$ with $k$ for
different values of $l$ with $a=0.2$ and $b=0.2$.}
\end{figure}

\begin{figure}[H]
\centering
\begin{subfigure}{ .58\textwidth}
\includegraphics[width=.8\linewidth]{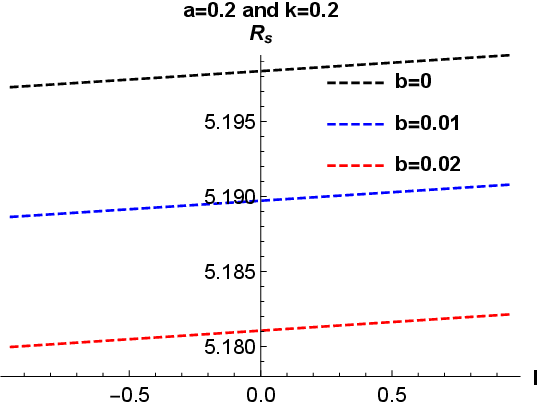}
\end{subfigure}%
\begin{subfigure}{ .58\textwidth}
\includegraphics[width=.8\linewidth]{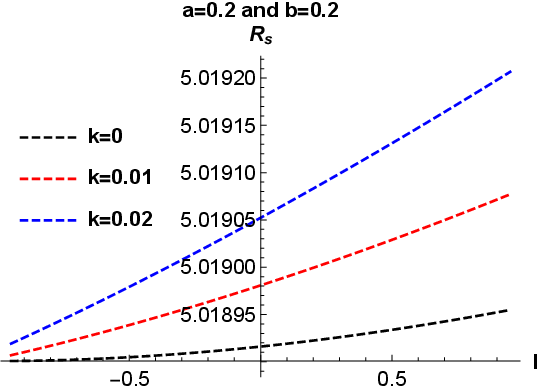}
\end{subfigure}
\caption{The  figure at the left side represent the variation of
$R_{s}$ with $l$ for different values of $b$ with $a=0.2$ and
$k=0.2$. The  figure at the right side represents the variation of
$R_{s}$ with $l$ for different values of $l$ with $a=0.2$ and
$b=0.2$.}
\end{figure}

\begin{figure}[H]
\centering
\begin{subfigure}{ .58\textwidth}
\includegraphics[width=.8\linewidth]{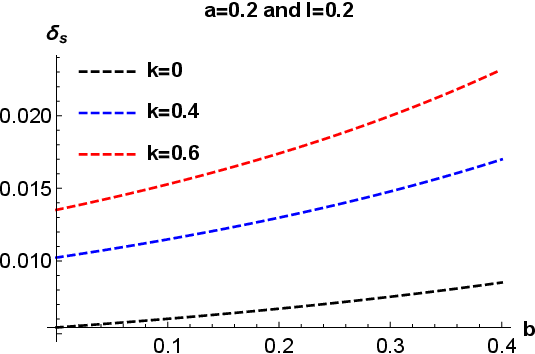}
\end{subfigure}%
\begin{subfigure}{ .58\textwidth}
\includegraphics[width=.8\linewidth]{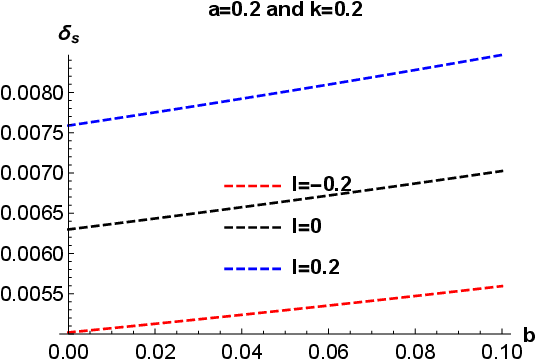}
\end{subfigure}
\caption{The figure at the left side represent the variation of
deviation from circularity $\delta_{s}$ with $b$ for different
values of $k$ with $a=0.2$ and $l=0.2$. The  figure at the right
side represents the variation of $\delta_{s}$ with $b$ for
different values of $l$ with $a=0.2$ and $k=0.2$.}
\end{figure}

\begin{figure}[H]
\centering
\begin{subfigure}{ .58\textwidth}
\includegraphics[width=.8\linewidth]{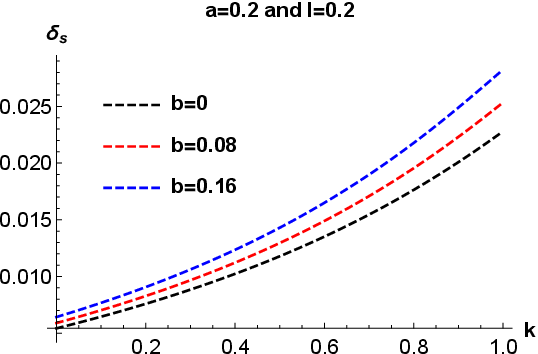}
\end{subfigure}%
\begin{subfigure}{ .58\textwidth}
\includegraphics[width=.8\linewidth]{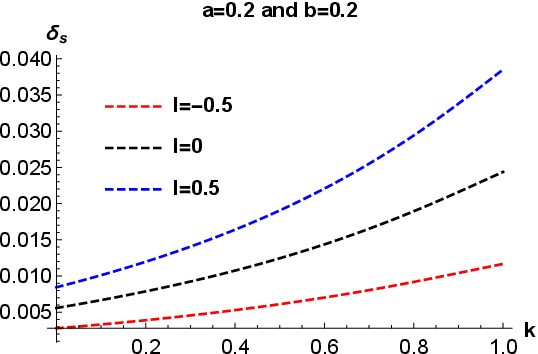}
\end{subfigure}
\caption{The  figure at the left side represents the variation of
deviation from circularity $\delta_{s}$ with $k$ for different
values of $b$ with $a=0.2$ and $l=0.2$. The  figure at the right
side represents the variation of $\delta_{s}$ with $k$ for
different values of $l$ with $a=0.2$ and $b=0.2$.}
\end{figure}

\begin{figure}[H]
\centering
\begin{subfigure}{ .58\textwidth}
\includegraphics[width=.8\linewidth]{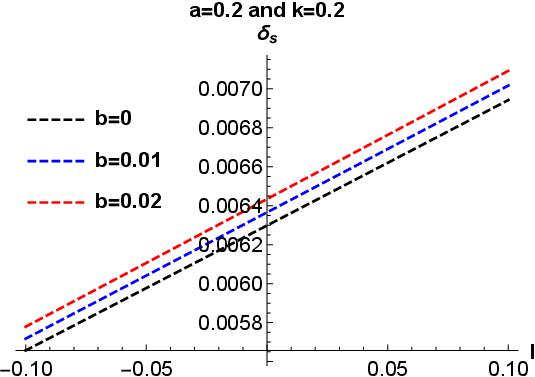}
\end{subfigure}%
\begin{subfigure}{ .58\textwidth}
\includegraphics[width=.8\linewidth]{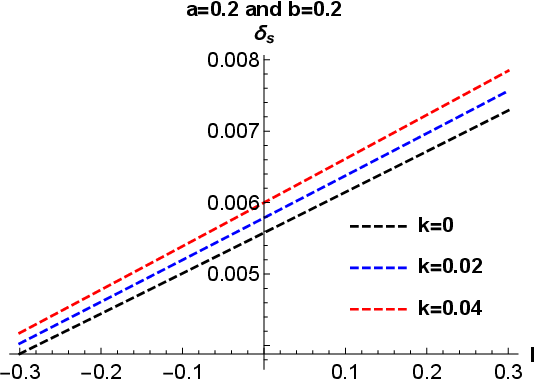}
\end{subfigure}
\caption{The  figure at the left side represents the variation of
deviation from circularity $\delta_{s}$ with $l$ for different
values of $b$ with $a=0.2$ and $k=0.2$. The  figure at the right
side represents the variation of $\delta_{s}$ with $l$ for
different values of $l$ with $a=0.2$ and $b=0.2$.}
\end{figure}
Here, we have given in a tabular form the deviation of the image
$\delta_{s}$ of the Kerr-Sen-like black hole for set of values of
$k \equiv \{0, .01, .02\}$ corresponding to the set of values of
$k\equiv \{-.2, 0, .2\}$ with $a=b=0.2$ and
$\theta=\frac{\pi}{2}$.
\begin{table}[h!]
\begin{center}
\caption{Deviation $\delta_{s}$ for Kerr-Sen-like black hole with
$a=0.2$ and $b=0.2.$ }
\begin{tabular}{|c|c|c|c|}
\hline
\textbf{} & \textbf{$l=-0.2$} & \textbf{$l=0$} & \textbf{$l=0.2$} \\
\hline
\textbf{$k=0$} & $0.00444613$ &$ 0.00557717$ & $0.00671623$ \\
\hline
\textbf{$k=0.01$} & $0.00452873 $& $0.00568118$ & $0.00684196$ \\
\hline
\textbf{$k=0.02$} & $0.00461217 $& $0.00578625$ & $0.00696899$ \\
\hline
\end{tabular}
\end{center}
\end{table}

\section{THE RATE OF ENERGY EMISSION}
It is known that black holes emit radiation and consequently,
the mass of the black hole decreases and the process continues
until it collapses down completely \cite{SW}. For various black
holes, this emission rate has been studied in the articles
\cite{WEI, GHOSH, GHOSH1}. In \cite{OUR}, we have studied the
energy emission rate for Kerr-Sen-like black holes and studied
the consequences of Lorentz violation effect $l$ on it. Here we
would like to study the energy emission in presence of plasma
following the motivation from the article \cite{GHOSH1}. The energy
emission rate of radiation with the frequency $\omega$ is given by
\begin{equation}
\frac{d^{2} E}{d \omega d t}= \frac{2 \pi^{3}
R_{s}^{2}}{e^{\frac{\omega}{T}}-1} \omega^{3}.
\label{EMRATE}
\end{equation}
where $\omega$ is the frequency and $T$ is the Hawking
temperature given by \cite{OUR}
\begin{eqnarray}
T=\frac{\sqrt{\left(2M-b\right)^2-4a^2\left(1+l\right)}} {4\pi M
\sqrt{1+l}\left(2M-b+\sqrt{\left(2M-b\right)^2
-4a^2\left(1+l\right)}\right)} \label{HT}.
\end{eqnarray}
In \cite{WEI}, it was conjectured that the black hole shadow
corresponds to its high energy absorption cross-section for the
observer located at infinity. In general, the absorption cross-section oscillates around a limiting constant value $\sigma_{lim}
= \pi R^2_s $ for a spherically symmetric black hole \cite{WEI,
MASH, MISNER}. The reader may see the articles \cite{MASH, MISNER}
where it was shown that $\sigma_{lim}$ was proportional to the
geometrical cross-section of its photon sphere. The fluctuation
around the limiting value was also studied in \cite{FLUC}. In
\cite{GHOSH}, it has been applied for studying energy emission. In
\cite{GHOSH1}, it was extended for the black hole in presence of
plasma. Therefore, the energy emission rate is directly proportional to
the surface area of the black hole, which may be considered to be
approximately equal to the surface area of the shadow $S\approx
\pi R_s^2$. Computing $R_s$ using the expression (\ref{RS}) and
using (\ref{HT}) we can calculate the rate of emission of
radiation, which enables us to plot energy emission versus
frequency of radiation curve. In Fig. 16, we have studied how the
energy emission rate will behave with the variation of the
parameters $b$, $l$, and $k$. It is observed that the rate of
emission is higher for smaller values of $b$. However, the reverse
is the case when $k$ increases. It is evident from the spectrum
that in the absence of plasma the minimum energy will be released
from the black hole, which indeed agrees with the Kerr-Newman
black hole \cite{BABAR}. From the plots, we see that for negative
$l$ there is an enhancement of emission whereas, for increasing
positive value of $l$, there is a reduction in the emission of radiation.

\begin{figure}[H]
\centering
\begin{subfigure}{ .58\textwidth}
\includegraphics[width=.8\linewidth]{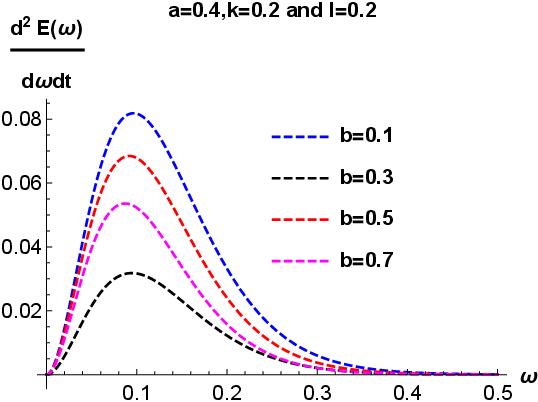}
\caption{Variation of the rate of emission with $\omega$ for\\
different values of $b$ with $a=0.4, k=0.2$, and $l=0.2$.}
\end{subfigure}%
\begin{subfigure}{ .58\textwidth}
\includegraphics[width=.8\linewidth]{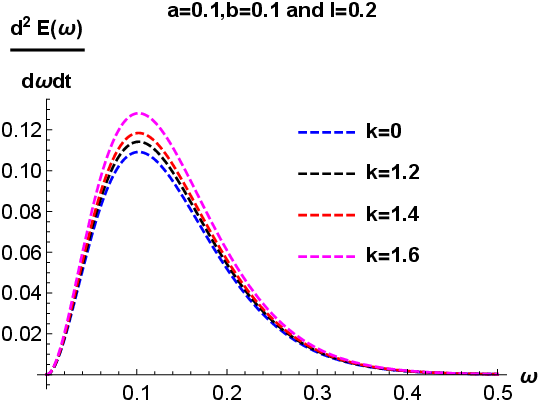}
\caption{Variation of the rate of emission with $\omega$ for\\
different values of $k$ with $a=0.1, b=0.1$, and $l=0.2$.}
\end{subfigure}
\par\bigskip
\begin{subfigure}{ .58\textwidth}
\includegraphics[width=.8\linewidth]{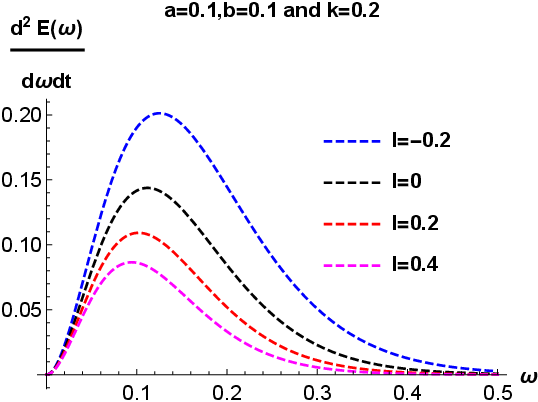}
\caption{Variation of the rate of emission with $\omega$ for
\\different values of $l$ with $a=0.1, b=0.1$, and $k=0.2$.}
\end{subfigure}
\caption{Variation of the rate of emission with $\omega$ for
different situations.}
\end{figure}

\section{Angle of Deflection of light in weak field in the presence of plasma}
The deflection of light or lensing in presence and in the absece
of dispersive medium like plasma for different spacetime
backgrounds has been studied in the earlier literature \cite{OY1,
OY2, OY3, CARL, MFAT, GKV}. The extension of it for this
Kerr-Sen-like black hole is worth investigation. There is an LV
parameter $l$. So from our study, the combined effect of the
presence of plasma and Lorentz violation on the deflection of
light can be estimated. To study it, weak field approximation will
be useful. This approximation is given by the expression
\begin{equation}
g_{\mu \nu}=\eta_{\mu \nu}+h_{\mu \nu},
\end{equation}
where $\eta_{\mu \nu}$ is the Minkowski metric and $h_{\mu \nu}$
is the perturbation metric over $\eta_{\mu \nu}$. The Minkowski
metric $\eta_{\mu\nu}= diag(-1,1,1,1)$ and $h_{\mu \nu}$ has
the the following properties under weak field approximation
\begin{eqnarray}\nonumber
h_{\mu \nu} \ll 1, \quad h_{\mu \nu} \rightarrow 0 \quad under \quad x^{i} \rightarrow \infty,\\
g^{\mu \nu}=\eta^{\mu \nu}-h^{\mu \nu}, \quad h^{\mu \nu}=h_{\mu \nu}.
\end{eqnarray}
Considering the weak plasma strength, the angle of deflection of
photon propagating along $z$ direction under weak field approximation is given by \cite{WEAK}
\begin{equation}
\hat{\alpha}_{k}=\frac{1}{2}
\int_{-\infty}^{\infty}\left(h_{33}+\frac{h_{00} \omega^{2}-K_{e}
N}{\omega^{2}-\omega_{e}^{2}}\right)_{, k} d z, \label{DEFLECTION}
\end{equation}
where $K_{e}=\frac{4\pi e^{2}}{m}$.
For large r, the black hole metric can, approximately, be written down
in the form
\begin{equation}
d s^{2}=d s_{0}^{2}+\left(\frac{2 M}{r}-\frac{2Mb}{r^{2}}\right) d t^{2}+\left(1+l\right)\left(\frac{2 M}{r}-\frac{2Mb}{r^{2}}\right) d r^{2},
\end{equation}
where
\begin{equation}
d s_{0}^{2}=-d t^{2}+\left(1+l\right)d r^{2}+r^{2}\left(d \theta^{2}+\sin ^{2} \theta d \phi^{2}\right).
\end{equation}
The components $h_{\alpha \beta}$ in the Cartesian coordinates
have the following expressions
\begin{eqnarray}\nonumber
h_{00}&=&\left(\frac{R_{g}}{r}-\frac{2Mb}{r^{2}}\right), \\\nonumber
h_{i k}&=&\left(1+l\right)\left(\frac{R_{g}}{r}-\frac{2Mb}{r^{2}}\right) n_{i} n_{k}, \\\nonumber
h_{33}&=&\left(1+l\right)\left(\frac{R_{g}}{r}-\frac{2Mb}{r^{2}}\right) \cos ^{2} \chi,
\end{eqnarray}
where $R_{g}=2 M$ and $\chi$ is the polar angle between 3-vector
and z-axis. Using the above expressions in (\ref{DEFLECTION}), the
light deflection angle for a black hole sitting in plasma medium
\cite{OY1, OY2, OY3, VS} is given by
\begin{equation}
\hat{\alpha}_{p}=\frac{1}{2} \int_{-\infty}^{\infty}
\frac{p}{r}\left(\frac{d h_{33}}{d r}+\frac{\omega^{2}}{\omega^{2}-\omega_{e}^{2}}
\frac{d h_{00}}{d r}-\frac{K_{e}}{\omega^{2}-\omega_{e}^{2}} \frac{d N}{d r}\right) d z,
\end{equation}
where $p^{2}=x_{1}^{2}+x_{2}^{2}$ refers to the impact parameter and
$x_{1}$ and $x_{2}$ are the coordinates on the plane orthogonal to
the $z$ axis. The expression of $r$ is $r=\sqrt{p^{2}+z^{2}}$. We
have the relation for the frequency of photon
\begin{equation}
\omega^{2}=\frac{\omega_{\infty}^{2}}{\left(1-\frac{R_{g}}{r}+\frac{2Mb}{r^{2}}\right)}.
\end{equation}
Here $\omega_{\infty}$ is the asymptotic value of photon
frequency. After expanding in series on the powers of
$\frac{1}{r}$, we can have the approximation
\begin{equation}
\left(1-\frac{\omega_{e}^{2}}{\omega^{2}}\right)^{-1} \simeq 1+\frac{4 \pi e^{2} N_{0} r_{0}}{m \omega_{\infty}^{2} r}-\frac{4 \pi e^{2} N_{0} r_{0} R_{g}}{m \omega_{\infty}^{2} r^{2}}.
\end{equation}
Using this approximation, one can find the deflection angle
$\hat{\alpha}_{p}$ of the light around a black hole in presence of
plasma in a straightforward manner
\begin{eqnarray}
\hat{\alpha}_{p}&=&\left(1+l\right) \frac{2 R_{g}}{p}+ \frac{2 R_{g}}{p}\left(1+\frac{\pi^{2} e^{2} N_{0} r_{0}}{m \omega_{\infty}^{2} p}-\frac{4 \pi e^{2} N_{0} r_{0} R_{g}}{m \omega_{\infty}^{2} p^{2}}\right) \\
&&-\frac{2Mb}{4 p^{2}}\left(3 \pi+\frac{4 \pi e^{2} N_{0} r_{0}}
{m \omega_{\infty}^{2} p}\left(8-\frac{3 \pi
R_{g}}{p}\right)\right).
\end{eqnarray}
We get $\hat{\alpha}_{p}=(1+l)\frac{2 R_{q}}{ p}$ in the absence
of charge and plasma for the Schwarzschild black hole. The
dependence of the angle of deflection $\hat{\alpha}_{p}$ on the
impact parameter $p$ for various charge and plasma parameters are
demonstrated in the figures below where $F=\frac{4 \pi e^{2} N_{0}
r_{0}}{m \omega_{\infty}^{2}}$. In the left panel, it is found
that as the value of $b$ increases the angle of deflection
decreases and it is seen that $\hat{\alpha}_{p}$ is maximum when
$b=0.$ We also observe that the deflection angle
$\hat{\alpha}_{p}$ increases with the increase in the plasma
parameter (right panel). We also notice that the deviation of
photons is smaller when the plasma factor is removed from the
black hole background. It is also observed that the deflection
angle increases with an increase in the value of the Lorentz
violating parameter $l$.

\begin{figure}[H]
\centering
\begin{subfigure}{ .58\textwidth}
\includegraphics[width=.8\linewidth]{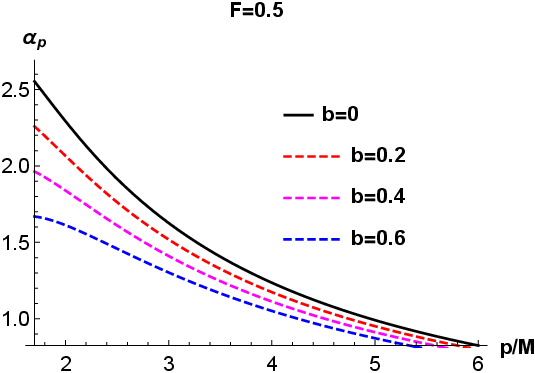}
\end{subfigure}%
\begin{subfigure}{ .58\textwidth}
\includegraphics[width=.8\linewidth]{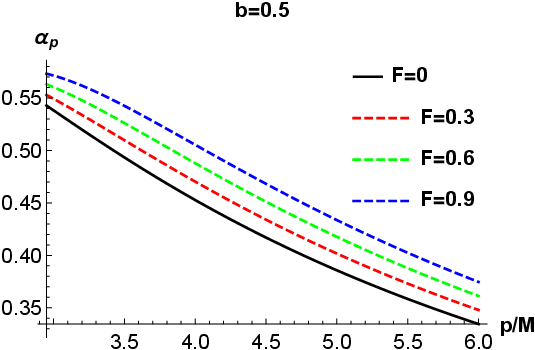}
\end{subfigure}
\par\bigskip
\begin{subfigure}{ .58\textwidth}
\includegraphics[width=.8\linewidth]{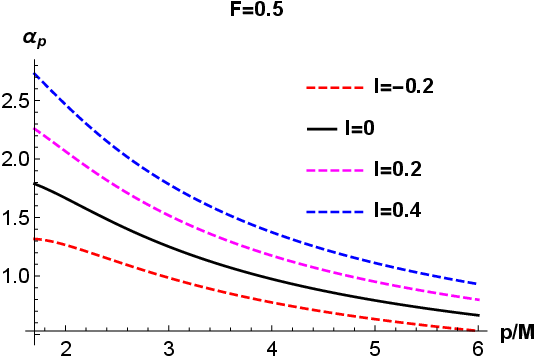}
\end{subfigure}
\caption{The plot at  upper left position  represents the
variation of angle of deflection $\hat{\alpha}_{p}$ with $p/M$ for
different values of $b$ with $F=\frac{4 \pi e^{2} N_{0} r_{0}}{m
\omega_{\infty}^{2}}=0.5$ and $l=0.2$. The plot at upper right
position describes the variation of $\hat{\alpha}_{p}$ with $p/M$
for different values of $F=\frac{4 \pi e^{2} N_{0} r_{0}}{m
\omega_{\infty}^{2}}$ with $b=0.5$ and $l=-0.4$. The lower one
represents the variation of $\hat{\alpha}_{p}$ with $p/M$ for
different values of $l$ with $F=\frac{4 \pi e^{2} N_{0} r_{0}}{m
\omega_{\infty}^{2}}=0.5$ and $b=0.2$.}
\end{figure}

\section{Constraining the parameter from the observed data for
$\mathrm{M}87^{*}$} We compare the shadows produced, from numerical
calculations, by the Kerr-Sen-like black holes with the observed one
for the $\mathrm{M}87^{*}$ black hole.  For comparison we consider the
experimentally obtained astronomical data for the circularity
deviation $\Delta \leq 0.10$ and angular diameter $\theta_{d}=42\pm 3 \mu as$ \cite{EHT1}. The boundary of the
shadow is described by the polar coordinate $(R(\phi),\phi)$ with
the origin at the center of the shadow $(\alpha_{C}, \beta_{C})$
where $\alpha_{C}=\frac{|\alpha_{max}+\alpha_{min}|}{2}$ and
$\beta_{C}=0$.

If a point $(\alpha, \beta)$ over the boundary of the image
subtends an angle $\phi$ on the $\alpha$ axis at the geometric
center, $\left(\alpha_{C}, 0\right)$ and $R(\phi)$ be the
distance between the point $(\alpha, \beta)$ and
$\left(\alpha_{C}, 0\right)$, then the average radius
$R_{\text{avg}}$ of the image is given by \cite{CBK}
\begin{equation}
R_{\text {avg}}^{2} \equiv \frac{1}{2 \pi} \int_{0}^{2 \pi} d \phi R^{2}(\phi), \\
\end{equation}
where $R(\phi) \equiv
\sqrt{\left(\alpha(\phi)-\alpha_{C}\right)^{2}+\beta(\phi)^{2}}$,
and $\phi = tan^{-1}\frac{\beta(\phi)}{\alpha(\phi)-\alpha_{C}}$.

With the above inputs, the deviation from circularity $\Delta C$
is defined by \cite{TJDP},
\begin{equation}
\Delta C \equiv 2\sqrt{\frac{1}{2 \pi} \int_{0}^{2 \pi} d
\phi\left(R(\phi)-R_{\text {avg }}\right)^{2}}.
\end{equation}
We also consider the shadow angular diameter which is define by
\begin{equation}
\theta_{d}=\frac{2}{d}\sqrt{\frac{A}{\pi}}
\end{equation}
where $A=2\int_{r_{-}}^{r_{+}} \beta d\alpha $ is the shadow area and $d=16.8 Mpc$
 is the distance of $M87^{*}$ from the earth. These relations will enable us to accomplish a comparison between
the theoretical predictions for Kerr-Sen-like black-hole shadows
and the experimental findings of the Event Horizon Telescope
collaboration.\\
Figures below represent the deviation from circularity
$\Delta C$ as it is obtained for Kerr-Sen-like black holes for the
angles of inclinations $\theta=90^{o}$ and $17^{o}$ respectively.

\begin{figure}[H]
\centering
\begin{subfigure}{.25\textwidth}
\centering
\includegraphics[scale=.7]{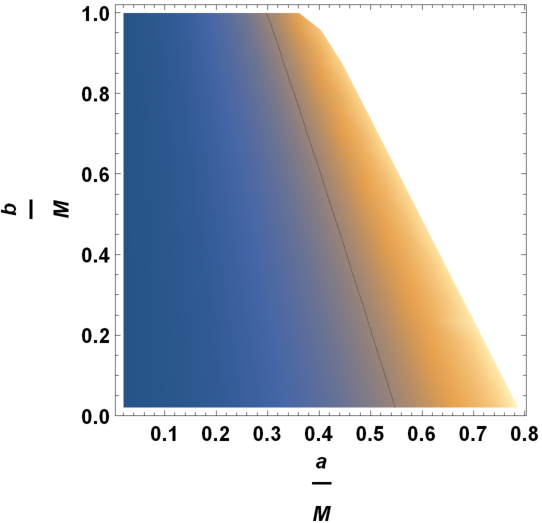}\hspace{1.5em}%
\end{subfigure}%
\begin{subfigure}{.28\textwidth}
\centering
\raisebox{.2\height}{\includegraphics[scale=.6]{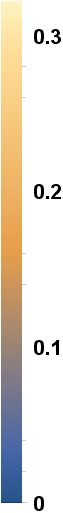}}
\hspace{1.5em}%
\end{subfigure}%
\begin{subfigure}{.25\textwidth}
\centering
\includegraphics[scale=.7]{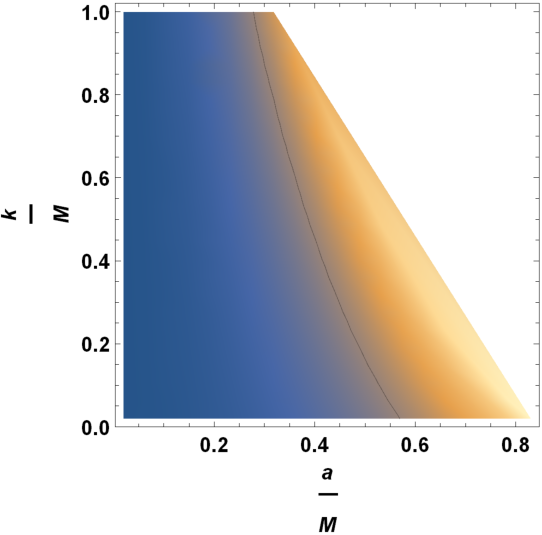}\hspace{1.5em}%
\end{subfigure}%
\begin{subfigure}{.28\textwidth}
\centering
\raisebox{.2\height}{\includegraphics[scale=.6]{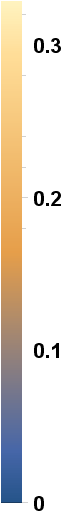}}
\end{subfigure}
\caption{The observable for deviation from circularity $\Delta C$
as a function of black hole parameters. The left one is for
$l=0.1$ and $k=0.1$. The right one is for $l=0.1$ and $b=0.1$. The
angle of inclination is $\theta= 90^{o}$. The black lines in the
figures correspond to $\Delta C=0.1$.}
\end{figure}
\smallskip

\begin{figure}[H]
\centering
\begin{subfigure}{.25\textwidth}
\centering
\includegraphics[scale=.7]{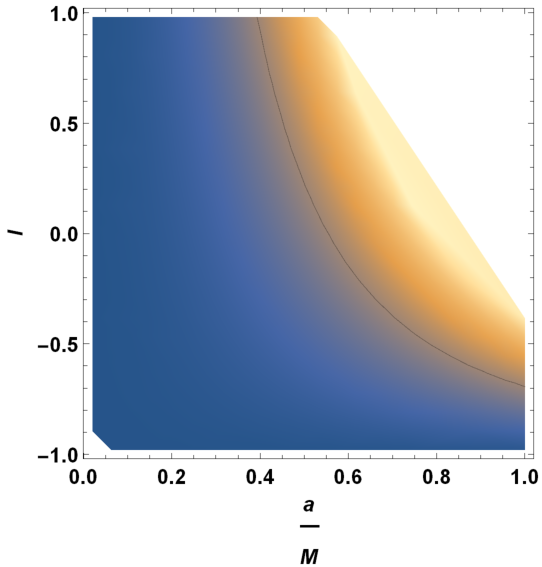}\hspace{1.5em}%
\end{subfigure}%
\begin{subfigure}{.28\textwidth}
\centering
\raisebox{.2\height}{ \includegraphics[scale=.6]{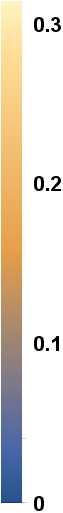}}
\hspace{1.4em}%
\end{subfigure}%
\begin{subfigure}{.25\textwidth}
\centering
\includegraphics[scale=.7]{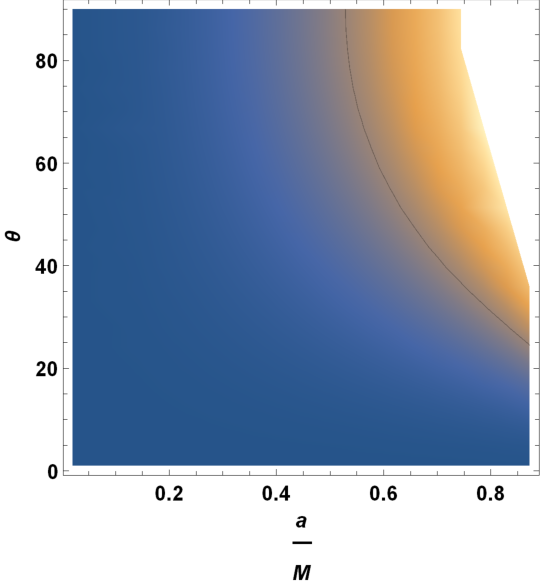}\hspace{1.5em}%
\end{subfigure}%
\begin{subfigure}{.28\textwidth}
\centering
\raisebox{.19\height}{\includegraphics[scale=.65]{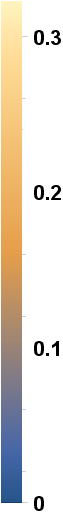}}
\end{subfigure}
\caption{The observable for deviation from circularity $\Delta C$
as a function of black hole parameters. The left one is for
$b=0.1$ and $k=0.1$. The right one is for $l=0.1$, $b=0.1$, and
$k=0.1$. Here the angle of inclination is $\theta= 90^{o}$. The
black lines in the figures correspond to $\Delta C=0.1$.}
\end{figure}

\begin{figure}[H]
\centering
\begin{subfigure}{.25\textwidth}
\centering
\includegraphics[scale=.7]{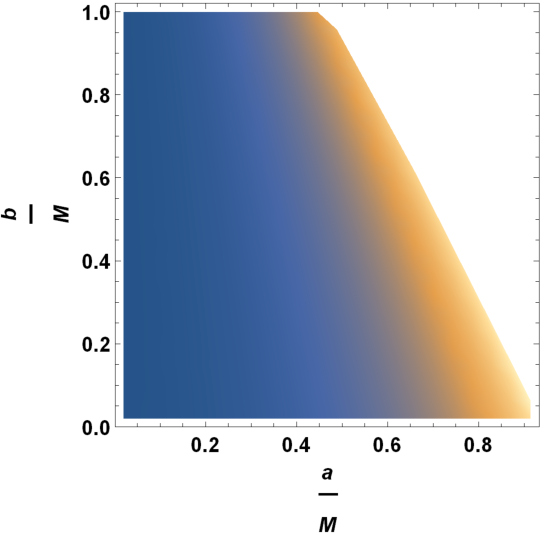}\hspace{1.5em}%
\end{subfigure}%
\begin{subfigure}{.28\textwidth}
\centering
\raisebox{.2\height}{ \includegraphics[scale=.6]{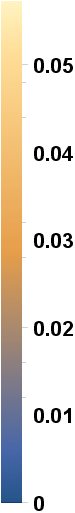}}
\hspace{1.5em}%
\end{subfigure}%
\begin{subfigure}{.25\textwidth}
\centering
\includegraphics[scale=.7]{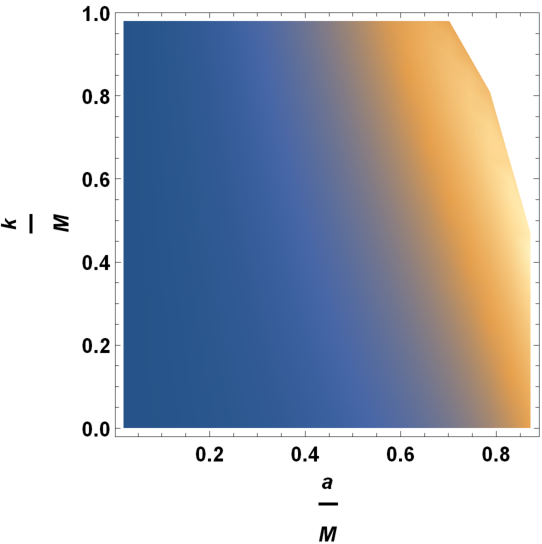}\hspace{1.51em}%
\end{subfigure}%
\begin{subfigure}{.28\textwidth}
\centering
\raisebox{.2\height}{\includegraphics[scale=.6]{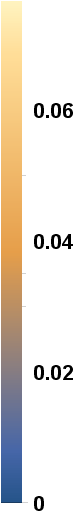}}
\end{subfigure}
\caption{The observable for deviation from circularity $\Delta C$
as a function of black hole parameters.
The left one is for $l=0.1$ and $k=0.1$. The right one is for
$l=0.1$ and $b=0.1$. Here the angle of inclination is $\theta=
17^{o}$.}
\end{figure}

\begin{figure}[H]
\centering
\begin{subfigure}{.25\textwidth}
\centering
\includegraphics[scale=.7]{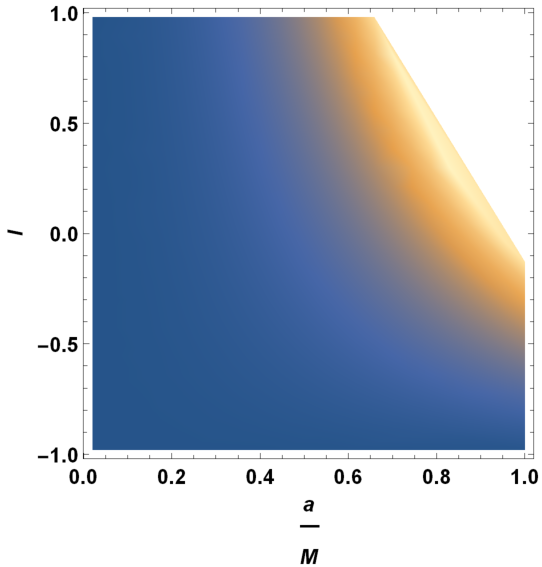}\hspace{1.5em}%
\end{subfigure}%
\begin{subfigure}{.28\textwidth}
\centering
\raisebox{.2\height}{ \includegraphics[scale=.6]{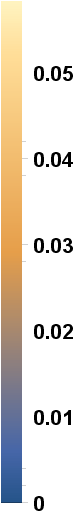}}
\hspace{1.5em}%
\end{subfigure}%
\caption{The observable for deviation from circularity $\Delta C$
as a function of black hole parameters. This is for $b=0.1$ and
$k=0.1$ with the angle of inclination $\theta = 17^{o}$.}
\end{figure}
Figures below represent the angular diameter $\theta_{d}$ as it is obtained for Kerr-Sen-like black holes for the
angles of inclinations $\theta=90^{o}$ and $17^{o}$ respectively.

\begin{figure}[H]
\centering
\begin{subfigure}{.25\textwidth}
\centering
\includegraphics[scale=.7]{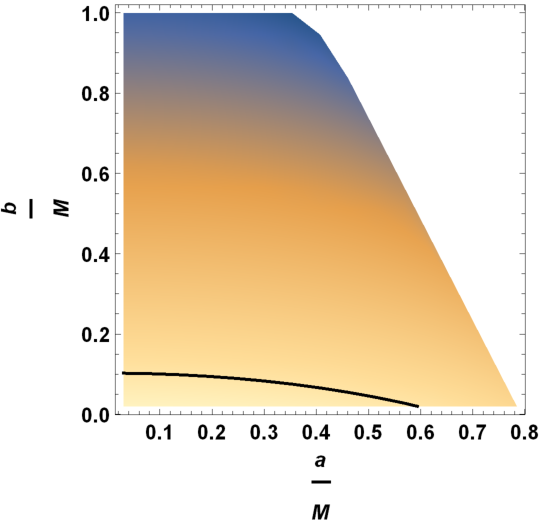}\hspace{1.5em}%
\end{subfigure}%
\begin{subfigure}{.28\textwidth}
\centering
\raisebox{.2\height}{ \includegraphics[scale=.6]{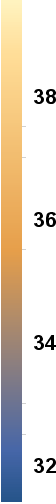}}
\hspace{1.5em}%
\end{subfigure}%
\begin{subfigure}{.25\textwidth}
\centering
\includegraphics[scale=.7]{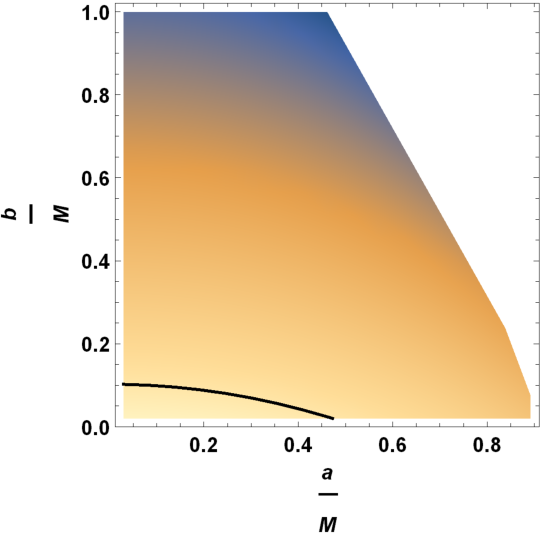}\hspace{1.51em}%
\end{subfigure}%
\begin{subfigure}{.28\textwidth}
\centering
\raisebox{.2\height}{\includegraphics[scale=.6]{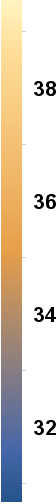}}
\end{subfigure}
\caption{The observable for angular diameter $\theta_{d}$
as a function of black hole parameters.
The left one is for $l=0.1$, $k=0.1$, and angle of inclination $\theta=90^{o}$. The right one is for
$l=0.1$, $k=0.1$, and $\theta=17^{o}$. Black solid lines correspond to $\theta_{d}=39 \mu as$}
\end{figure}

\begin{figure}[H]
\centering
\begin{subfigure}{.25\textwidth}
\centering
\includegraphics[scale=.7]{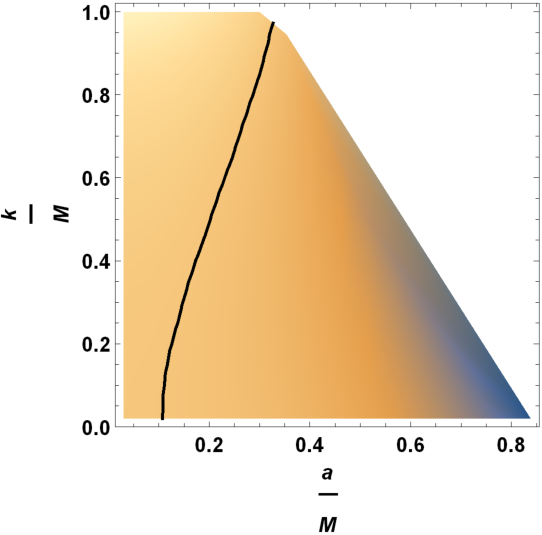}\hspace{1.5em}%
\end{subfigure}%
\begin{subfigure}{.28\textwidth}
\centering
\raisebox{.2\height}{ \includegraphics[scale=.6]{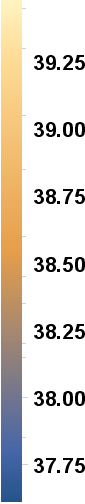}}
\hspace{1.5em}%
\end{subfigure}%
\begin{subfigure}{.25\textwidth}
\centering
\includegraphics[scale=.7]{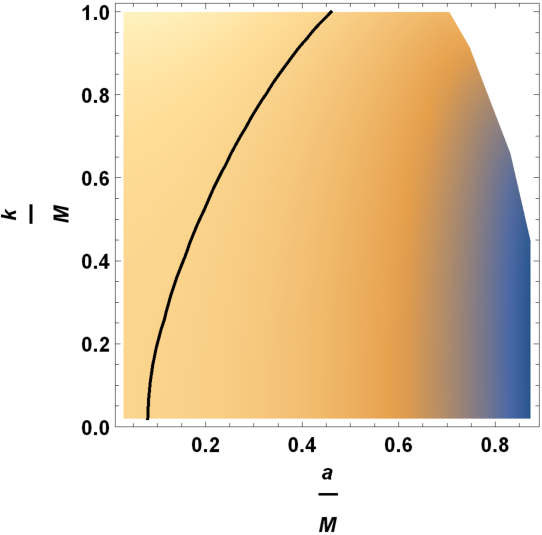}\hspace{1.51em}%
\end{subfigure}%
\begin{subfigure}{.28\textwidth}
\centering
\raisebox{.2\height}{\includegraphics[scale=.6]{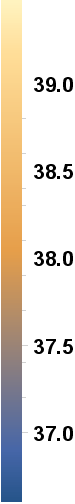}}
\end{subfigure}
\caption{The observable for angular diameter $\theta_{d}$
as a function of black hole parameters.
The left one is for $l=0.1$, $b=0.1$, and angle of inclination $\theta=90^{o}$. The right one is for
$l=0.1$, $b=0.1$, and $\theta=17^{o}$. Black solid lines correspond to $\theta_{d}=39 \mu as$}
\end{figure}

\begin{figure}[H]
\centering
\begin{subfigure}{.25\textwidth}
\centering
\includegraphics[scale=.7]{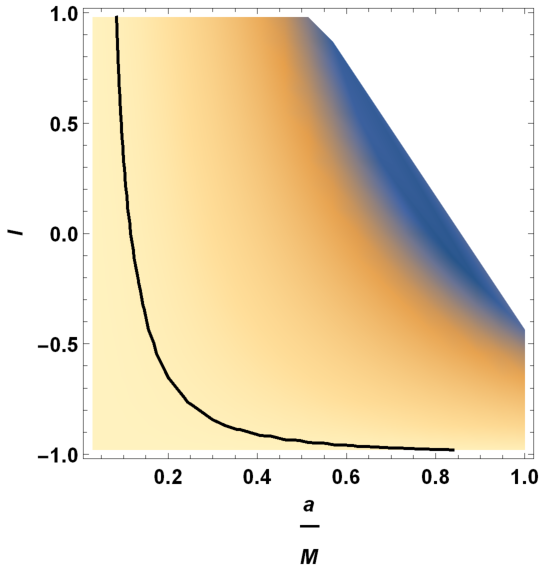}\hspace{1.5em}%
\end{subfigure}%
\begin{subfigure}{.28\textwidth}
\centering
\raisebox{.2\height}{ \includegraphics[scale=.6]{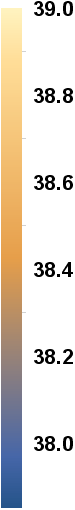}}
\hspace{1.5em}%
\end{subfigure}%
\begin{subfigure}{.25\textwidth}
\centering
\includegraphics[scale=.7]{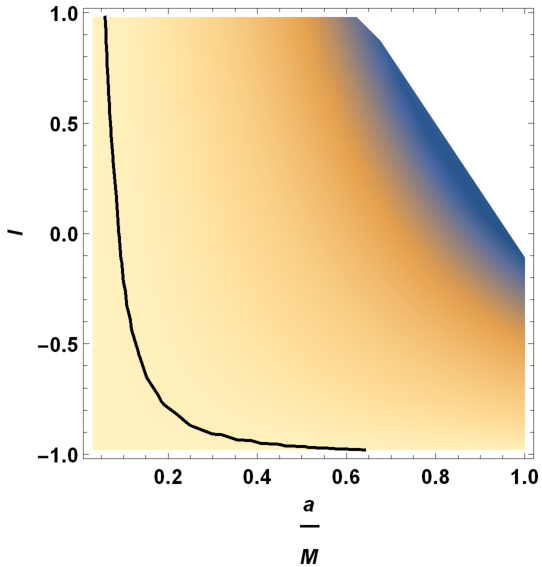}\hspace{1.51em}%
\end{subfigure}%
\begin{subfigure}{.28\textwidth}
\centering
\raisebox{.2\height}{\includegraphics[scale=.6]{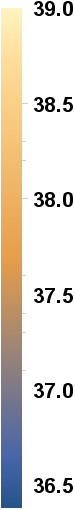}}
\end{subfigure}
\caption{The observable for angular diameter $\theta_{d}$
as a function of black hole parameters.
The left one is for $k=0.1$, $b=0.1$, and angle of inclination $\theta=90^{o}$. The right one is for
$k=0.1$, $b=0.1$, and $\theta=17^{o}$. Black solid lines correspond to $\theta_{d}=39 \mu as$}
\end{figure}

\begin{table}[h!]
\begin{center}
\caption{Deviation $\Delta C$ of Kerr-Sen-like black hole with $k=l=0.1$.
The left one is for the inclination angle $\theta=90^{o}$ and the right one
is for the inclination angle $\theta=17^{o}$. }
\begin{tabular}{|c|c|c|}
\hline
\text{a} & \text{b} & \textbf{$\Delta C$} \\
\hline
$ 0.020 $&$ 0.020 $&$ 0.000108539 $\\
\hline
$0.020 $&$ 0.265$ &$0.000135858$ \\
\hline
$0.020 $&$ 0.510 $& $0.000175677 $\\
\hline
$0.020 $&$ 0.755$ &$0.000237773$ \\
\hline
$0.020$&$ 1.000 $& $0.000343416 $\\
\hline
$0.265$&$ 0.020 $& $0.019818100 $\\
\hline
$0.265$ &$ 0.265$ &$ 0.025141500$ \\
\hline
$0.265$ &$ 0.510 $& $0.033223300 $\\
\hline
$0.265$ &$ 0.755 $&$0.046671500$ \\
\hline
$0.265$ &$ 1.000 $& $0.072912500$\\
\hline
$0.510 $& $0.020 $&$ 0.085970200$ \\
\hline
$0.510 $& $0.265$ &$ 0.112077000$ \\
\hline
$0.510 $& $0.510 $& $ 0.165146000$\\
\hline
\end{tabular}
\quad
\begin{tabular}{|c|c|c|}
\hline
\text{a} & \text{b} & \textbf{$\Delta C$} \\
\hline
$0.020 $&$ 0.020 $&$0.0000106680$\\
\hline
$ 0.020 $& $0.265$ &$0.0000132214$ \\
\hline
$ 0.020 $& $0.510 $& $ 0.0000154633 $\\
\hline
$ 0.020 $& $0.755$ &$ 0.0000203415$\\
\hline
$ 0.020 $& $1.000 $& $0.0000294412 $\\
\hline
$ 0.265$ &$ 0.020 $&$ 0.0017091800 $\\
\hline
$ 0.265$ &$ 0.265$ &$ 0.0021694100 $\\
\hline
$ 0.265$ &$ 0.510 $& $0.0028655300 $\\
\hline
$ 0.265$ &$ 0.755$ &$0.0040104600 $\\
\hline
$ 0.265$ &$ 1.000 $&$0.0061582600$ \\
\hline
$ 0.510 $& $0.020 $& $ 0.0073562800 $\\
\hline
$ 0.510 $& $0.265$ &$ 0.0097972700 $\\
\hline
$ 0.510 $& $0.510 $& $ 0.0140271000$\\
\hline
\end{tabular}
\end{center}
\end{table}
For this type of estimation, a prerequisite independent input is
the knowledge of the black hole mass, which is the intrinsic
yardstick for this system. In the case of $\mathrm{M}87^*$, the
mass has been estimated to be $M = (6.6 \pm 0.4)\times 10^{9}
M_{\odot}$ \cite{MASS}. Considering $M=6.5 \times 10^{9}
M_{\odot}$ for the $\mathrm{M} 87^{*}$ black hole, we use the
observed value of deviation from circularity $\Delta C \leq 0.10$
as deduced by the EHT collaboration \cite{EHT1} in order to
constrain the Kerr-Sen-like black hole parameters. We have made an
attempt to put constraints on $(a, k)$, $(a, b)$ and $(a,l)$ by
fixing the values of ($b,l$), ($k,l$) and ($b,k$) respectively.
From Fig. 18 and Fig. 19, it is clear that over a finite parameter
space, black hole shadows satisfy the bound $\Delta C \leq 0.10$.
Fig. 20 and Fig. 21 reveal that the condition $\Delta C \leq
0.10$ is satisfied by the Kerr-Sen-like black hole in presence of
plasma for the entire parameter space. From Fig. 22, Fig. 23 and
Fig.24 it is clear that, for finite parameter space, angular diameters
are within $1\sigma$ region i.e $\theta_{d}=39 \mu as$. From figures above
 we can also conclude that for fixed values of black hole
parameters, the apparent shadow circularity deviation and angular diameter
decrease as the observer moves away from the equatorial plane toward the polar axis.

We may put constrain on the value of $l$ as well. By
 modelling M$87^{*}$ black hole as Kerr black hole,
the author has found a lower limit a $a$ for the M$87^{*}$ black
hole \cite{RODRIGO}. If we bring this result under consideration
in the present situation the interval of interest for $a$ becomes
$[0.50M,0.99M]$ with $b=k=0$. Now if we combine the constraints
$\Delta C \leq 0.10$ and $\theta_{d}=42\pm3 \mu as$ and the
knowledge that $a\in[0.50M,0.99M]$, we observe that
$l\in(-1,0.621031]$. Here we find upper bound of $l$ to be
0.621031,  however $l$ cannot be equal to $-1$. So $l > -1$. So to
consider both the positive and negative value of $l$ in every plot
make sense and well justified

\section{Conclusion}
We have considered the Kerr-Sen-like black hole, which is a
solution of Einstein-bumblebee gravity. It contains an LV parameter
$l$. We study the propagation of light in a non-magnetized
pressure-less plasma on this Kerr-Sen-like spacetime. We have
considered the plasma as a medium with dispersive properties
given by a frequency-dependent index of refraction. The
gravitational field is determined by the mass, the spin, and the
bumblebee parameter. The Gravitational field due to plasma is
neglected here. We have studied how the nature of shadows gets affected
by the LV effect associated with the bumblebee field in the
pressure-less nonmagnetic plasma. We have also studied the energy
emission scenario and the weak field lensing in the Kerr-Sen-like
spacetime due to Einstein-bumblebee gravity background when it
is veiled in a plasma medium. The results for Kerr, Kerr-like, and
the Schwarzschild black holes can, as well, be obtained from this
result with a suitable limit. We have investigated in detail the
impact of the charge ($Q^2 \propto b$), plasma parameter $k$, and
the Lorentz violating bumblebee parameter $l$ on the structure of
shadow, on the lensing effect of light, and on the energy emission
due to black hole radiation. It has been found that the
new parameters quantitatively influence the structure of the event
horizon by altering the radius significantly. It is observed that the
size of the shadow viewed by a distant observer reduces with an
increase in the charge $Q$ and the size of the shadow appears to
be larger with the increase in the plasma parameter $k$ in the
presence of a bumblebee field.

Therefore, the bumblebee field maintains its action of deforming
the shadow in presence of plasma. Since the energy liberated
from the black hole depends on the area of the shadow, the rate of
energy emission from the black hole is higher when the black hole
is surrounded by a plasma. As far as the angle of deflection is
concerned, the photons are observed to experience an increase in
the deviation as the plasma factor $k$ increases. While on the
other hand, the angle of deflection reduces sufficiently when the
amount of charge parameter increases. It is also observed that the
deflection angle increases with an increase in the value of the LV
parameter $l$. Sufficient negative $l$ may cease the lensing
effect too!

The theory of General Relativity predicts that the Kerr metric is
capable of describing astrophysical black holes. However, such
black holes may not exist in a perfect vacuum due to the presence
of surrounding matter, like plasma, dark matter, etc. The
accretion disks may also alter the black hole, which may not fit
with the Kerr metric. The bumblebee modified Kerr spacetime
enables the incorporation of potential deviations from the Kerr
metric through the additional deviation parameters. The advantage
of this type of modified gravity is that it is a generic one and
it has the ability to cover the prediction of the Kerr metric
since the introduced parameters can be chosen in such a way that
it makes the null deviation of the shadow from Kerr spacetime.
There is the theoretical motivation for modified theories of
gravity indeed.

The EHT collaboration has captured the image of supermassive black
hole $\mathrm{M}87^{*}$ exhibiting a deviation from circularity
$\Delta C\leq 0.10$. The observation did not say anything about
modified theories of gravity or suggest any alternatives to the
Kerr black hole. Motivated by this, we considered Kerr-Sen-like
modified black holes, which have additional deviation parameters
$l$ and $b$ to the Kerr black hole. The presence of Plasma adds an
additional parameter. We use the deviation from circularity
$\Delta C \leq 0.10$ as deduced by the EHT collaboration to
constrain the Kerr-Sen-like black hole parameters. We have
observed that over a finite parameter space, black hole shadows
satisfy the bound $\Delta C \leq 0.10$ both in the presence of plasma
and in the absence of it.

Besides the result of the article \cite{RODRIGO} allows us to put
put constrain on the value of $l$. We have found that the value of
the parameter $l>-1$ and the upper bound of it is 0.621031. So $l$
can take both  positive and negative values.

Finally, we would like to mention that in the calculation of the
rate of energy emission we did not consider the greybody factor.
The mathematical formulation of it is much involved for any black
hole and it is important if a quantitative estimation is
attempted. Unfortunately, there is no available data to compare it
with the real situation. A theoretical evaluation of it for this
black hole may be done, which we are leaving for a separate study.

\end{document}